\RequirePackage{graphicx}
\documentclass[prd, twocolumn, superscriptaddress, nofootinbib, floatfix]{revtex4-2}

\usepackage{xcolor}
\usepackage[subfigure]{graphfig}
\usepackage[normalem]{ulem} 
\usepackage{wrapfig}
\usepackage{amsmath}
\usepackage{amsfonts}
\usepackage{amssymb}
\usepackage{booktabs}
\usepackage{multirow}
\usepackage{slashed}
\usepackage{physics}
\usepackage{tabularx}
\usepackage{soul}
\usepackage{ulem}
\usepackage{hhline}
\usepackage{float}
\usepackage[colorlinks=true, linkcolor=blue, citecolor=blue, urlcolor=blue]{hyperref}
\usepackage{longtable}

\usepackage{nccmath}
\usepackage{esvect}
\newcommand\befs{\begin{figure*}}
\newcommand\eefs[1]{\label{fig:#1}\end{figure*}}
\newcommand\bef{\begin{figure}}
\newcommand\eef[1]{\label{fig:#1}\end{figure}}
\newcommand\bet{\begin{table}}
\newcommand\eet[1]{\label{tb:#1}\end{table}}
\newcommand\bets{\begin{table*}}
\newcommand\eets[1]{\label{tb:#1}\end{table*}}

\newcommand{\bea}{\begin{eqnarray}}
\newcommand{\eea}{\end{eqnarray}}

\newcommand{\om}{{\omega}}
\newcommand{\nn}{\nonumber}

\newcommand{\Dl}{\Delta}

\newcommand{\ms}{\overline{\rm{MS}}}

\definecolor{amethyst}{rgb}{0.6, 0.4, 0.8}

\begin{document}

\title{Gluon helicity in the nucleon from lattice QCD and machine learning}

\newcommand*{\BRAC}{BRAC University, 66 Mohakhali, Dhaka 1212, Bangladesh}\affiliation{\BRAC}

\newcommand*{\SDU}{Key Laboratory of Particle Physics and Particle Irradiation (MOE), Institute of Frontier and Interdisciplinary Science, Shandong University, Qingdao, Shandong 266237, China}\affiliation{\SDU}

 \newcommand*{\UKY}{Department of Physics \& Astronomy, University of Kentucky, Lexington, KY 40506,
USA}\affiliation{\UKY}

\newcommand*{\RBRC}{RIKEN-BNL Research Center, Brookhaven National Laboratory, Upton, NY 11973, USA}\affiliation{\RBRC}

\newcommand*{\BNL}{Physics Department, Brookhaven National Laboratory, Upton, NY 11973, USA}\affiliation{\BNL}

\author{Tanjib~Khan}\email{ext.tanjib.atique@bracu.ac.bd}\affiliation{\BRAC}
\author{Tianbo~Liu}\email{liutb@sdu.edu.cn}\affiliation{\SDU}
\author{Raza~Sabbir~Sufian}\email{gluon2025@gmail.com}\affiliation{\UKY}\affiliation{\RBRC}\affiliation{\BNL}

\begin{abstract}
We present the first lattice QCD determination of  the light cone gluon helicity correlation parton distribution function (PDF) with numerical evidence toward disfavoring  negative gluon polarization in the nucleon. We present a  solution for eliminating an inevitable contamination term that dominates the Euclidean correlations and makes determining gluon helicity PDF unfeasible. The proposed synergy between lattice QCD and artificial intelligence  offers a superior platform to alleviate the defining challenge of extracting  quark and gluon PDFs from  the lattice data that are available in a limited domain due to a finite range of accessible hadron momenta. We suggest a systematically improvable method to extract PDFs from the lattice data, independent of inadequate parametrizations.  The result of the gluon helicity will improve our understanding of the role of spin in the strong interaction and the nucleon-spin structure.
\end{abstract}

\maketitle

 Understanding the internal structure and dynamics of protons and neutrons, which are complex many-body systems consisting of strongly interacting quarks and gluons, is at the core of exploring the visible matter in the Universe. Specifically, a profound knowledge of the origin of the proton's spin is critical for understanding the dynamics of the theory of strong interaction, quantum chromodynamics (QCD).  An ongoing  effort in theoretical and experimental  nuclear physics~\cite{EuropeanMuon:1987isl, EuropeanMuon:1989yki,Bunce:2000uv,PHENIX:2014gbf,HERMES:2008abz,COMPASS:2018pup}, including the future Electron-Ion Collider (EIC)~\cite{Accardi:2012qut,AbdulKhalek:2021gbh,Anderle:2021wcy}, is to realize the proton spin decomposition in terms of the quark and  gluon spin, and their orbital angular momenta~\cite{Jaffe:1989jz,Ji:1996ek} (for reviews, see~\cite{Aidala:2012mv,Leader:2013jra,Wakamatsu:2014zza,Deur:2018roz,Ji:2020ena,Liu:2021lke}).  An outstanding problem remains to explain why the quark spin contributes approximately 30\% to the proton spin, which is confirmed by analyzing experimental data~\cite{deFlorian:2009vb, Nocera:2014gqa, COMPASS:2015mhb, Ethier:2017zbq} and recent lattice QCD (LQCD) calculations~\cite{Lin:2018obj, Alexandrou:2020sml,Wang:2021vqy}. The challenge is to discern how much of the remaining spin budget is contributed by gluons.

In experiments,  the access to the spin-dependent gluon parton distribution function (PDF) or  the helicity PDF $\Dl g(x)$  in the nucleon  is mostly obtained from the kinematic region probed by the polarized proton-proton collisions at the Relativistic Heavy-Ion Collider (RHIC) in the momentum fraction region, $x\in [0.04,0.2]$. Excluding the extrapolation of $\Dl g(x)$ in the small-$x$ region, the global analyses~\cite{deFlorian:2014yva,Nocera:2014gqa} obtained sizable positive $\Dl g(x)$ and hence positive gluon polarization $\Dl G$ in the nucleon.  Depending on whether $\Dl g(x)$ is positive or negative or consistent with zero, the ratio of  the polarized to  the unpolarized gluon PDFs, $\Dl g(x)/ g(x)$ can be quite different in the entire $x$-domain~\cite{COMPASS:2015mhb}. Two recent analyses~\cite{Zhou:2022wzm,Whitehill:2022mpq} reported  that both  the positive and  the negative solutions of $\Dl g(x)$ were able to equally describe the experimental data.   Hence, determining the correct sign of the gluon helicity distribution is a crucial factor to comprehend  the proton spin structure.  While the future EIC aims to cover presently inaccessible kinematic regions  including the unexplored small-$x$ region~\cite{Aschenauer:2015ata,Borsa:2020lsz} and  provide a stringent constraint on $\Dl g(x)$, it is critically important to determine the nonperturbative $\Dl g(x)$ from  the first-principles LQCD calculation  as theoretical predictions to be tested  in existing and upcoming experimental programs. 

In recent years, several formalisms to determine  the $x$-dependent hadron structures from LQCD  have been proposed~\cite{Liu:1993cv,Detmold:2005gg,Braun:2007wv,Ji:2013dva, Ji:2014gla,Chambers:2017dov,Radyushkin:2017cyf,Ma:2014jla,Ma:2017pxb}.  For  reviews of these formalisms and recent lattice calculations see Refs.~\cite{Cichy:2018mum,Constantinou:2020hdm, Ji:2020ect,Constantinou:2022yye}. Among these, the  equal-time matrix elements of the bilocal operators composed of two gluon fields, known as the quasi-PDF  matrix elements~\cite{Ji:2013dva} can be used to determine $\Dl g(x)$ using the large momentum effective theory~\cite{Ji:2014gla}. It has been shown that various combinations of  the quasi-PDF gluon operators are multiplicatively renormalizable~\cite{Zhang:2018diq,Li:2018tpe}, as well as for the case of  the bilocal quark operators~\cite{Izubuchi:2018srq,Ji:2017oey,Green:2017xeu}. In this work, we use the pseudo-PDF approach~\cite{Radyushkin:2017cyf} based on the quasi-PDF  in the Fourier space and a coordinate-space factorization at small distances as proposed in~\cite{Braun:2007wv}.  The pseudo-PDF approach gives only the shape of the gluon helicity Ioffe-time distribution (ITD)~\cite{Gribov:1965hf,Ioffe:1969kf,Braun:1994jq} and a separate calculation of the gluon momentum fraction $\langle x \rangle_g$ is required for  a proper normalization. However, the multiplicative renormalizability of the quasi-PDF matrix elements  utilized for renormalization at short distances using a ratio method~\cite{Orginos:2017kos}  is simpler within this approach. In spite of  the recent progress in LQCD calculations of the unpolarized gluon PDF~\cite{Fan:2018dxu,Fan:2020cpa,HadStruc:2021wmh,Fan:2021bcr,Salas-Chavira:2021wui,Fan:2022kcb}, the determination of $\Dl g(x)$ has not yet been possible due to the presence of the nucleon boost, $p_z$-dependent contamination term in the off light cone  matrix elements~\cite{HadStruc:2022yaw}.   

 In this  letter, we propose a solution to this major problem by eliminating the $p_z$-dependent contamination. This enables the Euclidean correlation to be matched to the light cone correlation, which can be used to extract $\Dl g(x)$. We derive a mathematical relation and implement a machine learning algorithm to extract the correlation function, which is dominated by the leading-twist contribution. 

To determine $\Dl g(x)$ from LQCD, one needs to calculate matrix elements of  the gluon field $G_{\mu\nu}$ and its dual $\widetilde{G}_{\lambda\beta}=(1/2)\epsilon_{\lambda\beta\rho\gamma}G^{\rho\gamma}$ separated by a spatial Wilson line $W[z, 0]$~\cite{Ji:2013dva,Balitsky:2021cwr},
\begin{eqnarray}\label{eq1:ME}
\Dl M_{\mu\alpha;\lambda\beta}(z,p,s) &=& \bra{p,s}G_{\mu \alpha} (z) \, W[z, 0] \widetilde{G}_{\lambda \beta} (0) \ket{p,s}\nn \\
&& - (z \to -z),
\end{eqnarray}
where $z$ is the separation between the gluon fields,  $p$ is the nucleon four-momentum, and $s$ is the nucleon polarization. The combination which gives access to  the gluon helicity correlation with the least number of contamination terms is $ \Dl {\mathcal M}_{00}(z,p_z)\equiv \Dl M_{0i;0i}(z,p_z) + \Dl M_{ij;ij}(z,p_z)$; $i,j=x,y$ being perpendicular to the nucleon boost in the $z$-direction, $p= \{p_0, 0_\perp, p_z \}$~\cite{Balitsky:2021cwr}. Leveraging the multiplicative renormalizability of the  gauge link-related UV divergences  by forming the following ratio,
\bea \label{eq:rITDdef}
\Dl\mathfrak{{M}}(z,p_z)\equiv i \frac{[\Dl{\mathcal M}_{00}(z,p_z)/p_z { p_0}]/Z_{\rm L}(z/a_L)}{{\cal M}_{00}(z,p_z=0)/m_p^2}\, ,
\eea
we obtain the renormalization group invariant reduced pseudo-Ioffe-time distribution. Here, ${\cal M}_{00}(z,p_z) \equiv [M_{0i;i0}(z,p_z)+M_{ji;ij}(z,p_z)]$ is the spin-averaged matrix element related to the unpolarized gluon correlation~\cite{Balitsky:2019krf,HadStruc:2021wmh} and  the factor $1/Z_{\rm L} (z/a_L)$ is determined in~\cite{Balitsky:2021cwr} to cancel the UV logarithmic vertex anomalous dimension
of the $\Dl {\cal M}_{00}$ matrix element. As a function of  Lorentz invariant variables, $z^2$ and $\om \equiv zp_z$ (known as the Ioffe time~\cite{Braun:1994jq} or  the quasi light-front distance~\cite{Ji:2020brr}), $\Dl\mathfrak{{M}}$ can be expressed in terms of  invariant amplitudes, $\Dl\mathcal{M}_{sp}^{(+)}$ and $\Dl{\mathcal M}_{pp}$~\cite{Balitsky:2021cwr},
\begin{eqnarray} \label{eq:Ipform1}
\Dl \mathfrak{M}(\om,z^2) &=&  [\Dl {\mathcal M}_{sp}^{(+)}(\om,z^2) - \om \Dl{\mathcal M}_{pp}(\om,z^2) ]\nn \\
&& - \frac{m_p^2}{p_z^2} \om \Dl{\mathcal{M}}_{pp}(\om,z^2) \, .
\end{eqnarray}
In contrast, the light cone correlation that  gives access to $x\Dl g(x,\mu)$ at a scale $\mu$ is
\begin{eqnarray} \label{eq:lcITD}
\Dl{\mathcal{I}}_g(\om,\mu) &\equiv& i [\Dl{\mathcal{M}}_{sp}^{(+)}(\om,\mu) - \om \Dl{\mathcal{M}}_{pp}(\om,\mu)]\nn \\
&=&\frac{i}{2} \int_{-1}^{1} d x~ e^{-ix\om}x\Dl g(x,\mu),
\end{eqnarray}
and does not contain the  $m_p^2/p_z^2$ suppressed term as appeared in Eq.~\eqref{eq:Ipform1}.  A natural choice to suppress this contamination term is to calculate  $\Dl \mathfrak{M}(\om,z^2)$ at a very large momentum. However, even for the nucleon mass, $m_p=0.938\,{\rm GeV}$  and $p_z \approx 3~{\rm GeV}$, the suppression factor $m_p^2/p_z^2 \approx 0.1$ and the contamination term dominates the  matrix elements as $\om$  increases. In addition, achieving good signals for  the gluonic matrix elements at the physical point and $p_z>3$ GeV will be very challenging in the near future calculations.  It is therefore important to develop an approach to eliminate the contamination term in the limit of large $p_z$. This would allow the resulting matrix elements to be matched to the light-cone matrix elements of  the gluon helicity distribution.  The inability to remove this contamination and perform subsequent matching  has prevented the determination of the gluon helicity PDF from lattice QCD prior to this work. 

An alternative expression of $\Dl \mathfrak{M}(\om,z^2)$ shows  that this matrix element is  nonvanishing at $p_z=0$~\cite{HadStruc:2022yaw} and  the following subtraction 
\begin{eqnarray}
\label{eq:subrep}
&&\Dl {\mathfrak{M}}_{g,\,\rm sub}(\om,z^2) =  \Dl{\mathcal{M}}_{sp}^{(+)}(\om,z^2)  -\om  \Dl{\mathcal{M}}_{pp}(\om,z^2) \nn \\
 && \qquad -\om \frac{m_p^2}{p_z^2} [\Dl{\mathcal{M}}_{pp}(\om,z^2)-\Dl{\mathcal{M}}_{pp}(\om=0,z^2)],
\end{eqnarray}
removes the $\mathcal{O}(\om)$ contamination but  the residual higher-order contamination can become significant at large $\om$. 

\begin{figure}[htp]
	\centering
 	\includegraphics[width=3.4in, height=2.5in]{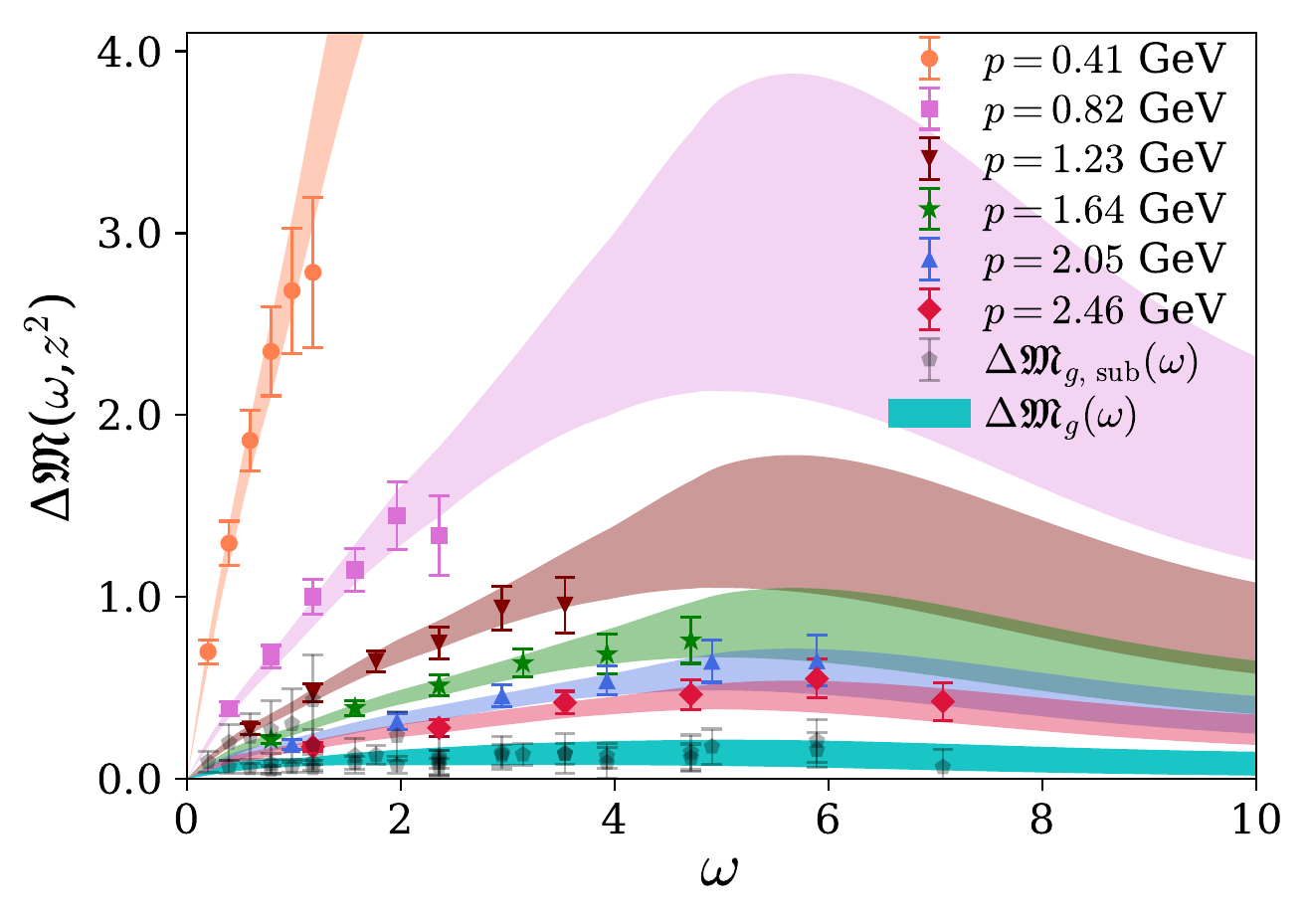}
 	\caption{Neural network fit to $\Dl {\mathfrak{M}}(\om)$ and $\Dl {\mathfrak{M}}_{g,\,\rm sub}(\om)$ gluonic matrix elements for all $p$ and $z_{\rm max}$ up to $6a$.  The cyan band represents the leading-twist dominated $\Dl {\mathfrak{M}_g}(\om)$.} \label{fig:NNpolITD}
 \end{figure}

In this work, we  first propose a solution to analytically eliminate the $(m_p^2/p_z^2)\om \Dl{\mathcal{M}}_{pp}$ contribution. We take advantage of the fact that different lattice boosts are related by $p_n=2\pi n/(La)$, where $a=0.094$ fm is the lattice spacing, and $L=32$ is the spatial extent of the lattice used in the calculation.  The technical details of the lattice QCD setups and the matrix elements that we use in this work can be found in~\cite{HadStruc:2022yaw}. For simplicity, we omit the subscript $z$ and write $p \equiv p_z$ in the rest of the paper and note that different lattice boosts $p_k$ and $p_l$ are related by the ratio  $r=p_k/p_l=k/l$ ($k>l$). Utilizing this relation and multiplying Eq.~\eqref{eq:Ipform1} by  the corresponding lattice squared-momentum $p_k^2$, we  obtain
\begin{eqnarray}
p_k^2\Dl \mathfrak{M}(\om)\big\vert_{p_k} &=&  p_k^2[\Dl {\mathcal M}_{sp}^{(+)}(\om) - \om \Dl{\mathcal M}_{pp}(\om) ]\nn \\
 &&- m_p^2 \om \Dl{\mathcal{M}}_{pp}(\om)\, .
\end{eqnarray}
and  another set of matrix elements by corresponding $p_l^2$, we   arrive at the following relation after subtraction:
\begin{eqnarray}
\label{eq:master}
\Dl \mathfrak{M}_g(\om) &\equiv& \Dl {\mathcal M}_{sp}^{(+)}(\om) - \om \Dl{\mathcal M}_{pp}(\om)\nn \\
&=& \frac{r^2 \Dl \mathfrak{M}(\om)\big\vert_{p_k}-\Dl \mathfrak{M}(\om)\big\vert_{p_l}}{r^2-1}\, .
\end{eqnarray}
Finally, $\Dl \mathfrak{M}_g(\om)$  is free of the contamination term and can be matched to  $\Dl{\mathcal{I}}_g(\om,\mu)$. The immediate challenge of implementing Eq.~\eqref{eq:master} is that the subtractions between multiple $p_n$ data sets require continuous functions in $\om$ and $\Dl \mathfrak{M}$  matrix elements at the same $\om$, but the LQCD data  in Fig.~\ref{fig:NNpolITD} are obtained at discrete $\om$ values and the ranges of $\om$ vary with $p_n$.  

To solve this problem of parametrization using moments and  to determine a contamination-free $\Dl \mathfrak{M}_g(\om)$, we perform a correlated simultaneous analysis to $\Dl \mathfrak{M}(\om,z^2)$ and $\Dl {\mathfrak{M}}_{g,\,\rm sub}(\om,z^2)$ data for all values of $p_n$  using neural network (NN), which is essentially a very flexible function parametrization. The imposed constraint in Eq.~\eqref{eq:master} serves as the main  assumption for the NN.  It is worth noting that the  NN alone cannot achieve the significant physics results presented in this paper. Instead, it is the simple but elegant Eq.~\eqref{eq:master} that we have derived in this study serves as a regulator for the  NN and allows us to obtain a nonzero and positive gluon helicity corrleation for the first time from a lattice QCD calculation as we demonstrate in the following.  

We parametrize $\Dl {\mathfrak M}$ and $\Dl {\mathfrak M}_{g,\,\rm sub}$ into $\Dl I$ plus power correction terms  according to Eqs.~\eqref{eq:Ipform1} and~\eqref{eq:subrep} as
\begin{align}
    \Delta {\mathfrak M}(\om) &= \Delta  I(\omega) + \frac{C(\omega)}{p_z^2},
    \label{eq:DM}\\
    \Delta {\mathfrak M}_{g,\rm sub}(\om) &= \Delta  I(\omega) + \frac{D(\omega)}{p_z^2},
    \label{eq:DMsub}
\end{align}
where $\Dl I(\omega)$, $C(\om)$, and $D(\om)$ are to be determined. The term $\Dl I(\omega)$ is further parametrized into a prefit function $\Dl I_0(\omega)$ multiplied by a deviation function $\delta {\cal I}(\omega)$.  In practice, to speed up the convergence, one can introduce a prefit function instead of directly parametrizing the goal function with  an NN.  Starting from a point closer to the solution can significantly improve the efficiency of the fitting though  being mathematically equivalent. Introducing the prefit function can accelerate the convergence to a smooth function in the fitting, while the result is not sensitive to any particular reasonable choice of $\Dl I_0(\omega)$. Here we take a fit curve from Ref.~\cite{Sufian:2020wcv}.

 \begin{figure}
    \centering
    \includegraphics[width=3.4in, height=2.2in]{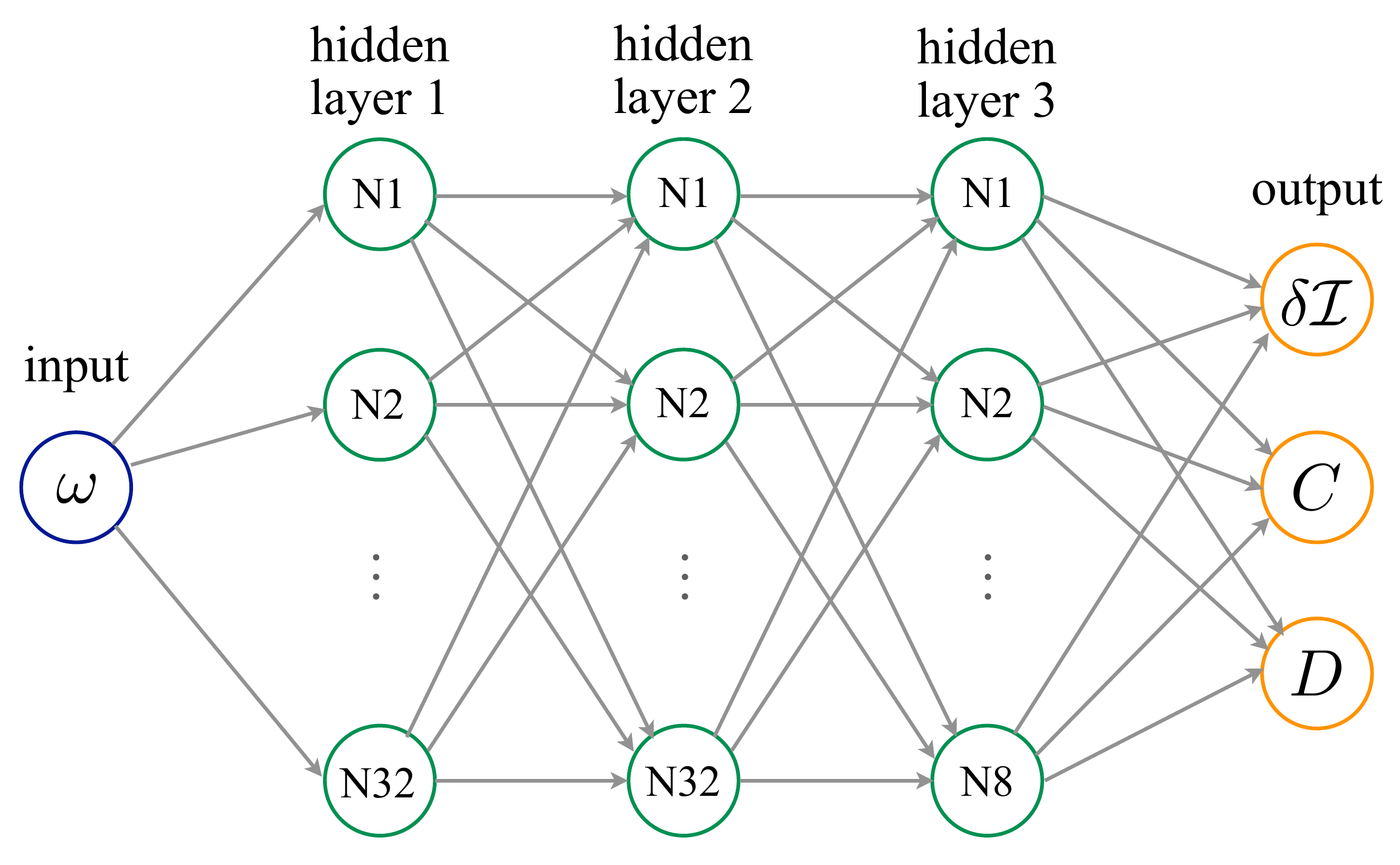}
    \caption{Architecture of the neural network. Each circular node represents a neuron and the arrows represent  the connection from the output of one artificial neuron to the input of another. }
    \label{fig:nn3}
\end{figure}

The architecture of the  NN for is illustrated in Fig.~\ref{fig:nn3}. The input layer contains one neuron for the $\omega$ value. The output layer contains three neurons for $\delta\cal{I}(\omega)$, $C(\omega)$ and $D(\omega)$ respectively. Three hidden layers, containing 32, 32, and 18 neurons respectively, are inserted between the input layer and the output layer. They are densely connected to  the corresponding former layers, as illustrated in Fig.~\ref{fig:nn3}, and activated with the rectified linear unit function. The output layer is densely connected to the last hidden layer and activated with the sigmoid function, which is renormalized and shifted to return values between $-20$ and $20$, a large enough range to cover any reasonable results.

The fitting procedure is to minimize the loss function. It is defined as the $\chi^2$ between the lattice data of $\Delta \mathfrak{M}$ and $\Delta \mathfrak{M}_{g,\rm sub}$ and the model values, which are calculated from the output values of $\delta\cal{I}(\omega)$, $C(\omega)$ and $D(\omega)$. There are in total 1901 paired sets of $\Delta {\mathfrak M}$ data and $\Delta {\mathfrak M}_{g,\rm sub}$. To capture the correlation between the unsubtracted data and the subtracted data, we pick at each time one set of $\Delta {\mathfrak M}$ and the corresponding $\Delta {\mathfrak M}_{g,\rm sub}$ to perform a simultaneous fit. In each fit, we randomly select 5/6 of the data from the paired set to create the training sample and leave the remaining 1/6 of the data in the validation sample. To keep the possibility  of finding multiple minima, the initial parameters of the NN are randomly generated. The loss value of the full data set is monitored during the training. It generally decreases at the beginning and starts to increase when overfitting happens, with small fluctuations from epoch to epoch all the time.  To prevent the accidental selection of small loss value points, some early epochs are eliminated. We stop the training process when there is no improvement in the total loss value for 3000 epochs and revert to the best result obtained. The result from the epoch with the least total loss function is saved. At the end, we have 1901 NNs corresponding to the 1901 paired set. The variation among these NNs reflects the uncertainty from the data.

 We also note that, for each of the particular choices of the network, we start the fitting from a randomly chosen point in the parameter space for the possibility  of finding different minima.  The results always converge to the same region. In addition, the partition of the training sample and the validation sample was not fixed  from time to time, and  the result does not show dependence on  the partition within the uncertainties inheriting from the lattice data.  As we also investigate below,  the systematic uncertainty from the choice of the NN is   negligible and the uncertainty is dominated by the lattice data.

 At this point, we discuss the advantage of using Eq.~\eqref{eq:master} and the NN analysis over conventional moments fits to remove the contamination term discussed above. In~\cite{HadStruc:2022yaw}, a fit to  the data using an expansion in moments was performed in an attempt to isolate $[\Dl {\mathcal M}_{sp}^{(+)}(\om) - \om \Dl{\mathcal M}_{pp}(\om)]$ and  to obtain a continuous distribution in $\om$ among different  $p_n$ data sets. An attempt to add only the second moment  to parametrize the contamination term containing $\Dl \mathcal{M}_{pp}(\om)$ resulted in an uncontrolled error and one needed to use the first moment as a Bayesian prior before the error in $\Dl \mathfrak{M}_g(\om)$ would blow up. We demonstrate this in the following. 
 
 In the expression of the gloun matrix elements
\begin{eqnarray} 
\Dl \mathfrak{M}(\om) &=&  [\Dl {\mathcal M}_{sp}^{(+)}(\om) - \om \Dl{\mathcal M}_{pp}(\om) ]\nn \\
&& - \frac{m_p^2}{p_z^2} \om \Dl{\mathcal{M}}_{pp}(\om) \, ,
\end{eqnarray}
$\Dl {\mathcal M}_{sp}^{(+)}$ is an odd function of $\om$ and $\Dl{\mathcal{M}}_{pp}$ is an even function of $\om$. Writing these amplitudes in terms of  the odd and even moments, one can parametrize the lattice data as
\bea \label{eq:parametrization}
\Dl \mathfrak{M}(\om) = \sum_{i=0} \frac{(-1)^i}{(2i+1)!}a_i \om^{2i+1} +  \om \frac{m_p^2}{p_z^2} \sum_{j=0} \frac{(-1)^j}{(2j)!}b_j \om^{2j} \ , \nn \\
\eea
where the coefficients $a_i$ are the Mellin moments of the gluon helicity reduced pseudodistribution related to $[ \Dl {\mathcal M}_{sp}^{(+)}(\om) -\om  \Dl{\mathcal{M}}_{pp}(\om)]$ and the $b_i$ are those for $\Dl{\mathcal{M}}_{pp}(\om)$.

\begin{figure}[htp]
	\centering
 	\includegraphics[width=3.4in, height=2.5in]{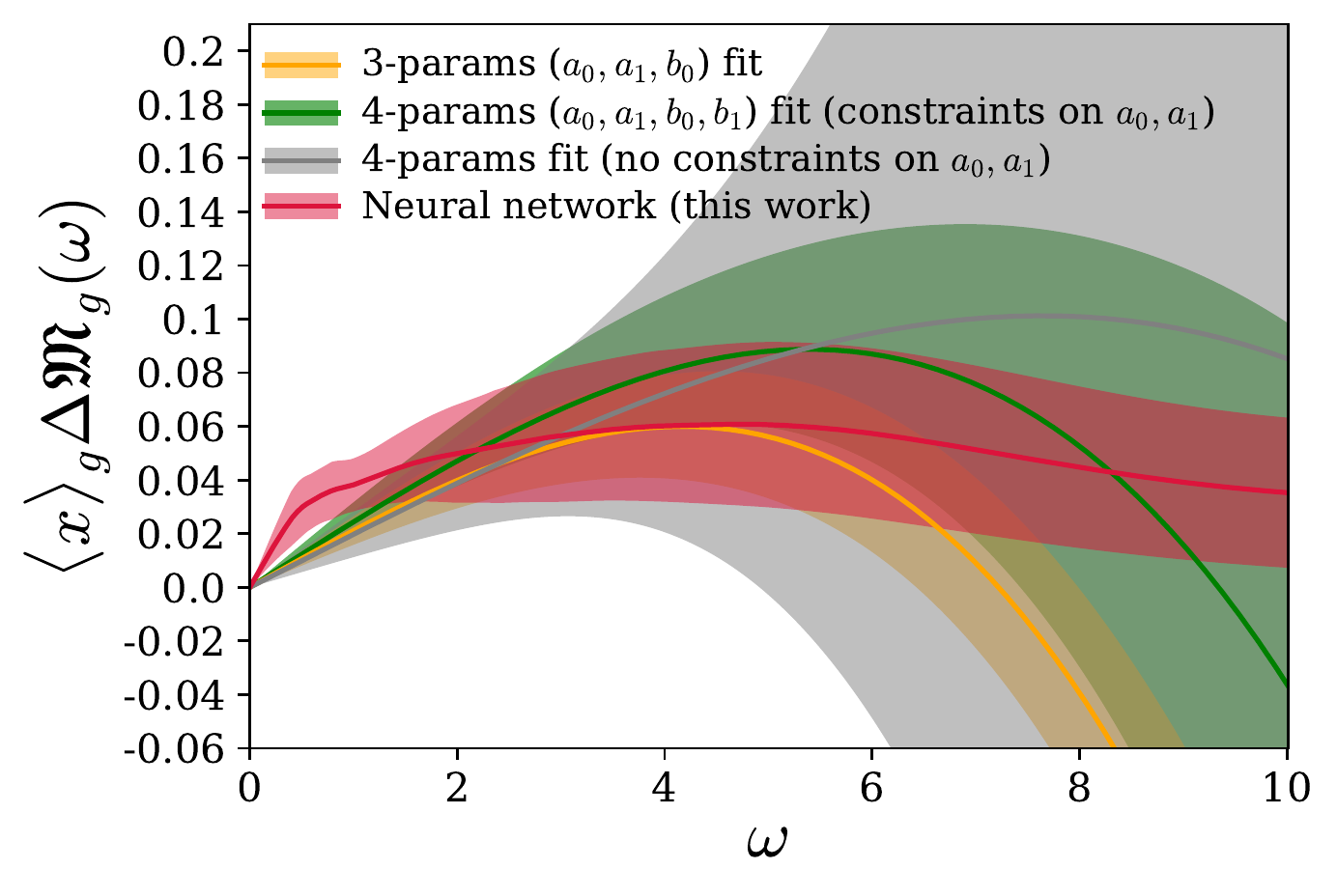}
 	\caption{Contamination term corrected matrix
element to determine $\Dl \mathfrak{M}_g(\om)$ associated with the gluon helicity distribution, normalized by the gluon momentum fraction. The orange band represents a fit with $i=0,1$ and $j=0$ in the parametrization Eq.~\eqref{eq:parametrization}. The green band represents fit with $i=0,1$ and $j=0,1$ by imposing $a_0,a_1, b_0$ from the previous fits as  the Bayesian priors. The gray band represents a fit with $i=0,1$ and $j=0,1$ but this time no Bayesian priors are imposed on the moments from the three-moment fit. The red band is the fit result from the neural network analysis. }\label{fig:comparison}
 \end{figure}

Fig.~\ref{fig:comparison} shows the extrapolated $\Dl \mathfrak{M}_g(\om)= \sum_{i=0} \frac{(-1)^i}{(2i+1)!}a_i \om^{2i+1}$ in the limit of zero ${\cal O}(m_p^2/p_z^2)$ contamination-term contribution within the fit parametrization using moments, normalized by the gluon momentum fraction $\langle x\rangle_g$.  The $\Dl \mathfrak{M}_g(\om)$ from the neural network analysis (NN) by imposing the constraint in Eq.~\eqref{eq:master}  is also shown in the figure for comparison.
%
 
 As shown in Fig.~\ref{fig:comparison},  instead of  $i=0,1$ in the expression for odd moments $\sum_{i} \frac{(-1)^i}{(2i+1)!}a_i \om^{2i+1}$, had it been used $i=0,1,2$, the $\Dl {\mathfrak M}_g(\om)$   fit in~\cite{HadStruc:2022yaw} would diverge upward as also have been demonstrated  in~\cite{Saalfeld:1997uv,Sufian:2020wcv}. The uncertainties and the downward  divergent trend of the fitted $\Dl {\mathfrak M}_g(\om)$  toward negative values  using moments depend on the truncation of the number of moments, rendering the fitting procedure unreliable and biased. Similar arguments go for a fit to the  noisier $\Dl {\mathfrak M}_{g,\rm sub}(\om)$ data using moments. As shown in Fig.~\ref{fig:comparison}, a three or four-parameters fit using Bayesian priors imposed on the fitted moments produce nonzero results of $\Dl \mathfrak{M}_g(\om)$ but invalidates the fit results  at  $\om \gtrsim 4$ due to model dependence. On the other hand, a $4$-parameter fit without Bayesian prior imposed on the first three moments produces huge uncertainty, and the $\Dl \mathfrak{M}_g(\om)$ is not usable for extracting  the PDF. In contrast, the NN with imposed constraint in Eq.~\eqref{eq:master} uses data up to $\om \approx 7$ and produces $\Dl \mathfrak{M}_g(\om)$ in an extended region beyond the LQCD data, clearly showing a significant advantage over model-dependent extractions.  Around $\om=11$, the NN extrapolation starts to show oscillation, which is expected further outside the lattice data and we consider $\Dl \mathfrak{M}_g(\om)$ up to $\om_{\rm max}=10$ in the subsequent analysis. 
 
 One may observe that the NN fit does not describe the  data well at $z>6a$.  This possibly indicates that the data points for $z=7a$ and $8a$ do not satisfy Eq.~\eqref{eq:master}, which is imposed to constrain the  outcome from the NN, and have significant higher-twist contributions. Therefore, they are not suitable for extracting the leading-twist dominated $\Dl \mathfrak{M}_g(\om)$.  We,  therefore, use data up to $z_{\rm max}=6a \approx 0.56$\,fm in our analysis. Although the set in of the higher-twist contribution can be observable dependent and data at $z \gtrsim 1~{\rm fm}$ with the assumption of the validity of short-distance factorization has been used in LQCD calculations, e.g.~\cite{Egerer:2021ymv},  a recent  calculation~\cite{Bhat:2022zrw} with the implementation of the 2-loop matching~\cite{Li:2020xml} found  that  the higher-twist contribution can become significant above $z\gtrsim 0.5$ fm (see for other findings~\cite{Karthik:2021sbj,Ji:2022ezo,Su:2022fiu}). With future precise gluonic matrix elements, it remains an open field for investigations  to determine  up to which $z_{\rm max}$ LQCD data is dominated by  the leading-twist contribution and  machine learning can be a useful tool for this study.

\begin{figure}[htp]
	\centering
 	\includegraphics[width=3.4in, height=2.5in]{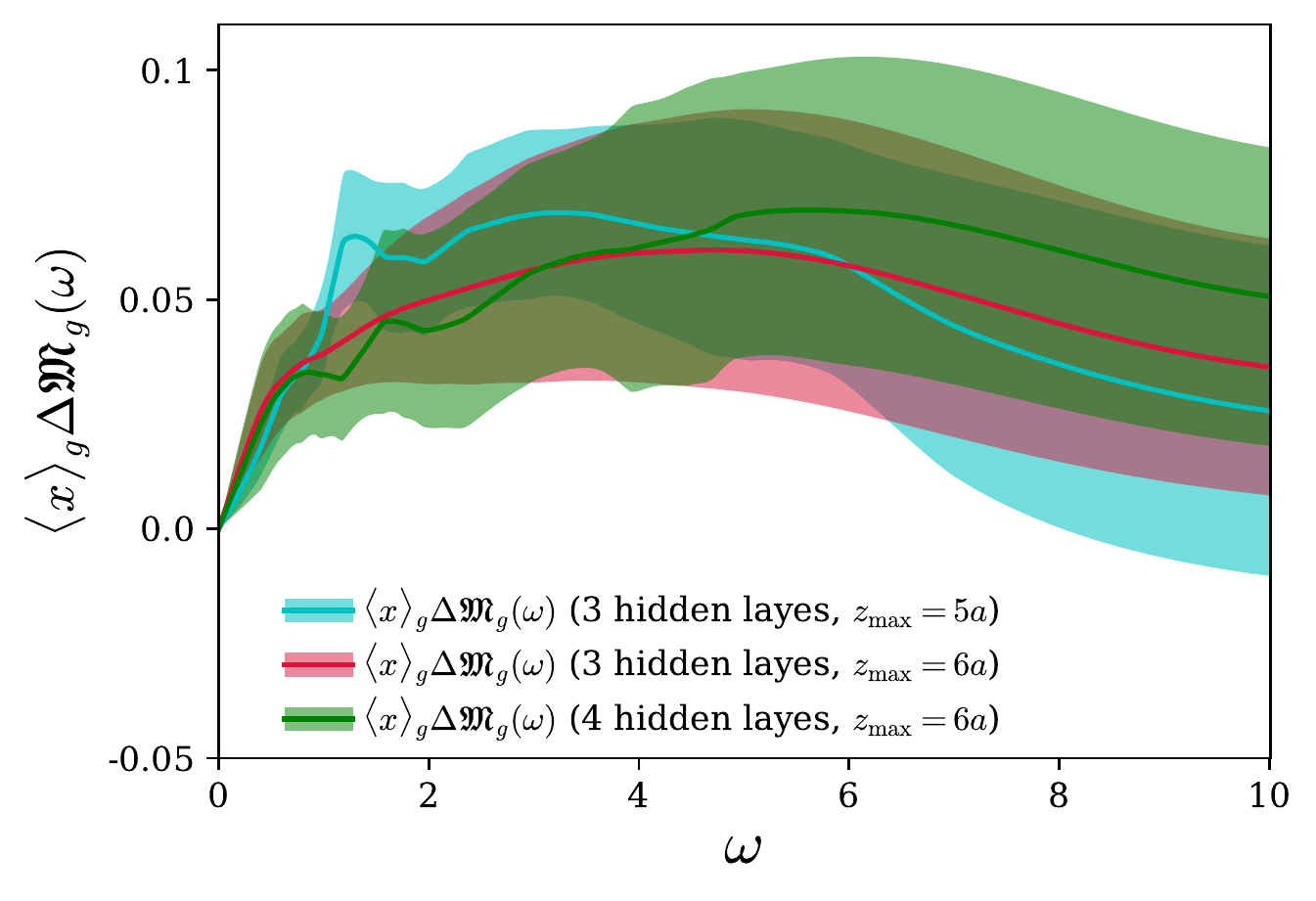}
 	\caption{Numerical investigations for the variation of $\langle x\rangle_g\Dl\mathfrak{M}_g(\om)$  due to the variation in the neural network analysis. The results  with  $z_{\rm max}=5a$ and $6a$, and the number of hidden layers used in this analysis are shown here.}\label{fig:NNlayers}
 \end{figure}

 Limiting to $z_{\rm max}=5a$ does not change the outcome of the NN analysis  significantly within  the uncertainty. To explore the systematic uncertainty from the choice of the particular NN, we repeat the analysis using  an NN with one additional hidden layer, {\it i.e.} an NN with four hidden layers containing 32, 32, 32, and 8 neurons respectively. More hidden layers mean a more flexible parametrization of the function, but as  shown in Fig.~\ref{fig:NNlayers}, the results of $\langle x\rangle_g \Delta {\mathfrak M}_g(\om)$ are consistent with each other within  the uncertainty.  We also  compare the results of the NN analysis by including and excluding the $z_{\rm max}=6a$ data points in the analysis. Since $z_{\rm max}=5a$ analysis leads to one less data point in the training sample for each momentum, it is expected to produce relatively larger  uncertainty compared to when $z_{\rm max}=6a$ data points are included in the analysis.

\begin{figure}[htp]
	\centering
 	\includegraphics[width=3.4in, height=2.5in]{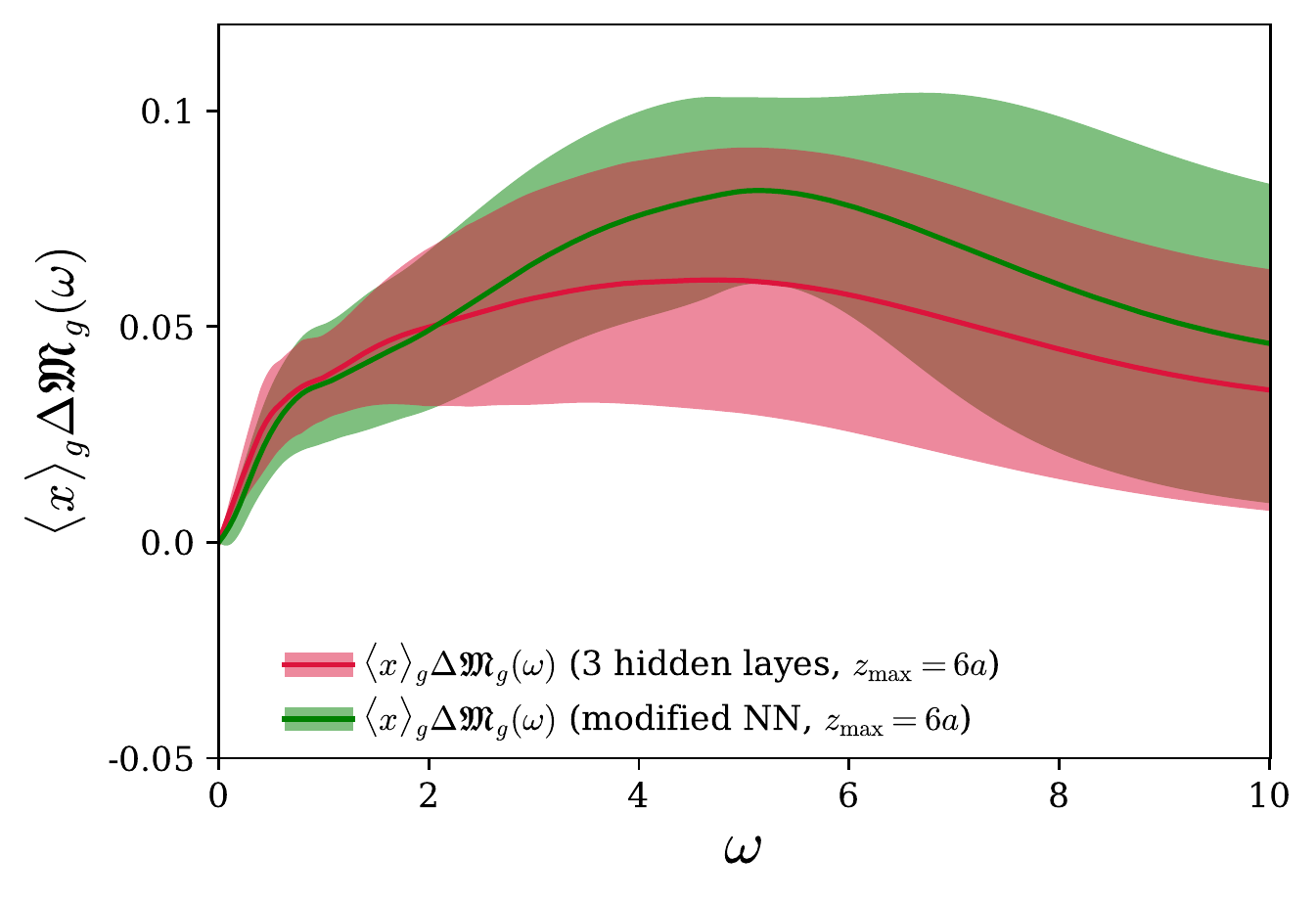}
 	\caption{Numerical investigations for the variation of $\langle x\rangle_g\Dl\mathfrak{M}_g(\om)$  due to the variation in the neural network analysis. The results are shown with  $z_{\rm max}=6a$. The NN analysis with four hidden layers and different outputs and activation functions is labeled as the modified neural network  (modified NN) as described in the main text. \label{fig:ML-4alter}}
 \end{figure}

  We also investigate the elimination of the contamination term with different networks and training structures and determine the $\langle x\rangle_g \Delta {\mathfrak M}_g(\om)$.  We choose a four-hidden-layer NN, which is demonstrated to be flexible enough.  We acknowledge that due to the limited lattice data, the neural network may not explore the full spectrum of potential architectures and solutions.  Nevertheless, it should strive for a level of generality beyond specific parametrizations, as long as the results remain consistent across variations in the number of layers, neural network architectures, and the choice of $z_{\rm max}$. As $\omega$ is the only feature in the parametrization, Eqs.~\eqref{eq:DM} and~\eqref{eq:DMsub}, it  is still taken for the input layer. The output layer contains three neurons, for the $\delta I$, $C/\Delta I$, and $D/\Delta I$ respectively. The hidden layers are activated by the exponential linear unit function, and the output layer is activated by the hyperbolic tangent function. We refer to this setup as the modified neural network or ``modified NN" in the subsequent texts and in Fig.~\ref{fig:ML-4alter}. We present a comparison plot between the $\langle x\rangle_g \Delta {\mathfrak M}_g(\om)$ determined with the initial neural network setup as shown in Fig.~\ref{fig:NNpolITD} and this modified setup in Fig.~\ref{fig:ML-4alter}.

  As we demonstrate in the above analyses, whether we change the number of layers in the NN, vary $z_{\rm max}$ between $5a$ and $6 a$, or change the setup of the NN, the results remain consistent  within uncertainties.  The most critical finding  from the analysis is that   our lattice QCD calculation favors  a positive $\langle x\rangle_g \Delta {\mathfrak M}_g(\om)$ in the available range of $\om$ remains unchanged. Therefore, the systematic uncertainty from the choice of the NN is negligible in comparison with the uncertainty of the extracted distribution,  reflecting that the uncertainty of the lattice data dominates  in the calculation. Naturally, machine learning methods cannot encompass every conceivable variation, but we  introduced variations in terms of layers, neural network architecture, and the upper limit parameter $z_{\rm max}$ in order to explore the resulting variances. It's important to highlight that there exists only one input, denoted as $\om$, and no variations are introduced at the input layer.

\begin{figure}[htp]
	\centering
 	\includegraphics[width=3.5in, height=2.1in]{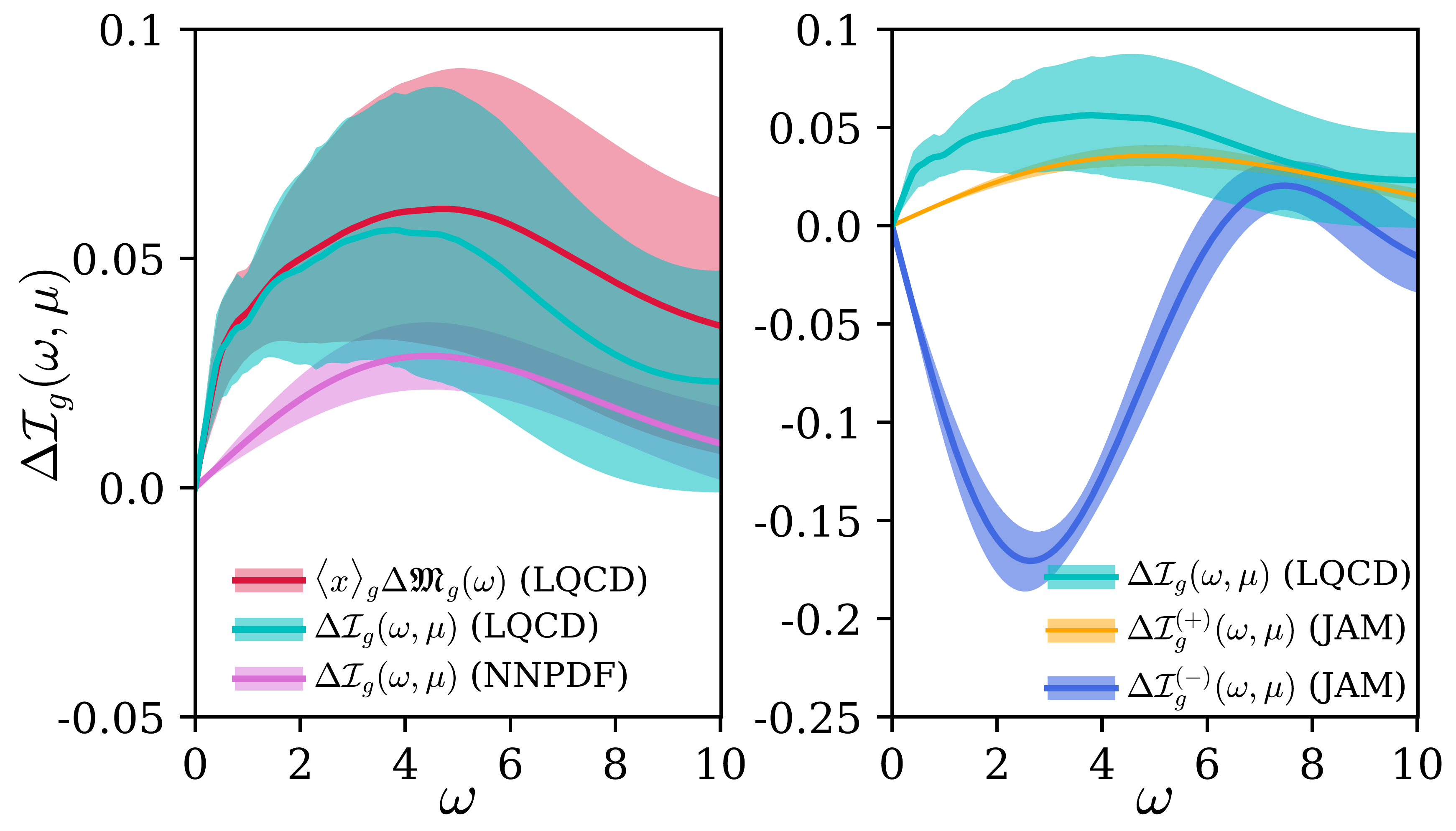}
 	\caption{Comparison of  the light cone gluon helicity ITD $\Dl{\mathcal I}_g(\om,\mu)$ with phenomenological results by the NNPDF~\cite{Nocera:2014gqa} (left panel) and  the JAM~\cite{Zhou:2022wzm} (right panel) Collaborations. The orange $\Dl{\mathcal{I}}_g^{(+)}$  and the blue  $\Dl{\mathcal{I}}_g^{(-)}$ bands represent the gluon helicity ITD  corresponding to the JAM positive and negative $x\Dl g(x)$ solutions, respectively. Pseudo-ITD, $\langle x\rangle_g\Dl{\mathfrak{M}_g} (\om)$ is  shown for comparison by the red band in the left-panel figure.} \label{fig:matchedITD}
 \end{figure}

The contamination-free $\Dl {\mathfrak{M}_g}(\om)$ can now be matched to the light cone  $\Dl{\mathcal I}_g (\om, \mu )$ and  the singlet quark ITD $\Dl{\mathcal{I}}_S (\om, \mu)$ in the $\ms$ scheme using the factorization relation~\cite{Balitsky:2021cwr} up to power corrections,

\begin{eqnarray} \label{eq:matching}
&&\Dl{\mathfrak{M}_g} ( \om ) \langle x \rangle_g(\mu) \!=\!    
\Dl{\mathcal I}_g (\om, \mu )\! -\!  \frac{\alpha_s N_c }{2\pi}\int_0^1 d u\,  \Dl{\mathcal I}_g (u\om, \mu )\nn \\
&&    \bigg\{ \ln\bigg(z^2 \mu^2 \frac{e^{2\gamma_E}}{4}\bigg)   \bigg( \bigg[\frac{2u^2} {\bar{u}} + 4u\bar{u}  \bigg]_+ - \bigg(\frac{1}{2}  + \frac{4}{3}  \frac{\langle x_S \rangle{(\mu)}}
 {  \langle x_g \rangle{(\mu)} } \bigg)  \nn \\
&&\delta( \bar{u} ) \bigg)+ 4 \bigg[\frac{u+\ln (1-u)}{\bar{u}}\bigg]_+ - \bigg( \frac{1} {\bar u} - \bar{u} \bigg)_+  -\frac{1}{2} \delta(\bar u)       \nn \\
&&+2\bar uu \bigg\}- \frac{ \alpha_s C_F}{2\pi}  \int_0^1 d u \,   \Dl{\mathcal{I}}_S (u \om,\mu )  
 \bigg\{\ln \bigg(z^2 \mu^2 \frac{ e^{2 \gamma_E}} {4 } \bigg)  \nn \\
 &&\Dl {\mathcal B}_{gq} (u)+ 2\bar uu    \bigg\}  \, ,
\end{eqnarray}
where $N_c=3$, $\bar{u} \equiv (1 - u)$, $\gamma_E$ is the Euler–Mascheroni constant, $\Dl {\mathcal B}_{gq}= 1-(1-u)^2$,  and   $\langle x \rangle_g(\mu=2~{\rm GeV})$ is chosen as $0.427$  from~\cite{Alexandrou:2020sml}.  Current statistics for  the gluonic matrix elements do not allow us to observe any logarithmic $z^2$-dependence and  thus it is not implemented  while extracting $\Dl \mathfrak{M}_g$.   We  choose $z=2a$ and  $\mu=2$ GeV in the matching Eq.~\eqref{eq:matching} and ignore the effect of  the singlet-quark contributions (which requires separate calculations). Taking the singlet distribution from  a global fit, e.g. NNPDF~\cite{NNPDF:2017mvq}, we find almost no observable impact on the matched $\Dl{\mathcal I}_g$. Similarly, varying values of $z$ or $\mu$ has minimal effects on the matched $\Dl{\mathcal I}_g$ within  the current statistical uncertainty.  This can be seen from the proximity of $\langle x \rangle_g \Dl \mathfrak{M}_g$ and $\Dl{\mathcal I}_g$ bands shown in Fig.~\ref{fig:matchedITD}.

From Fig.~\ref{fig:matchedITD}, it is evident that  the LQCD calculation of $\Dl{\mathcal I}_g$  disfavors the ITD constructed from  the negative $x\Dl g(x)$ solution from the global analysis in~\cite{Zhou:2022wzm}. Determination of ITDs from  the negative solution in~\cite{Zhou:2022wzm} with varying  the lower $x$-limits, for example,  $0.01\leq x \leq 0.99$ or $0.1 \leq x \leq 0.99$  can still be shown to be ruled out by our calculation in  the $\om \leq 10$ region. This is  the most important physics outcome of this LQCD calculation regarding the constraint on the large negative gluon helicity PDF in the moderate to large-$x$ values.

In principle,  the gluon helicity in the proton can be obtained from the ITD~\cite{Braun:1994jq}, $\Dl G(\mu) \equiv \int_{0}^{\infty} d \om~\Dl{\mathcal{I}}_g(\om,\mu)$. On the other hand, integrating  $\Dl \mathcal{I}_g(\om,\mu)$ up to $\om_{\rm max}=10$ from our calculation, we obtain 
\bea \label{eq:deltaG}
\Dl G^L(\mu) \equiv \int_{0}^{\om_{\rm max}} d \om~\Dl{\mathcal{I}}_g(\om,\mu)= 0.405(196)\, .
\eea
$\Dl G^L(\mu)$ is of course limited by the upper limit of the integration. While $\Dl G^L(\mu)$ in Eq.~\eqref{eq:deltaG} is a well-defined LQCD measure, the long-tail  of the ITD governed by the Regge behavior outside $\om_{\rm max}$  can lead to an underestimation (for a long positive tail) or an overestimation (for a negative tail) or some cancellations (for sign change in the tail) of this moment. An example of such an underestimation of the Gegenbauer moment in the pion distribution amplitude calculation can be seen in~\cite{Gao:2022vyh}. One can also try to extrapolate the ITD beyond $\om_{\rm max}=10$ using NN or perform phenomenological/model extrapolation of the Regge tail outside the LQCD data~\cite{Ji:2020brr,Gao:2021dbh} and get an estimate of the change in $\Dl G^L$.  However, the $\Dl G^L$ obtained here is expected to depend on  the pion mass, lattice spacing, and finite volume and we  refrain from extrapolating $\Dl G^L$. On the positive side, a phenomenological analysis~\cite{Sufian:2020wcv} found  that ITD in $\om \leq 6$  is the most affected region for different values of $\Dl G \in [0.2,0.4]$. The   $\Delta G^L(\mu)=0.405(196)(081)$, where the second uncertainty is the systematic uncertainty arising from variations of the neural network analysis, is about $3.8$-sigma away from the $\Delta G(\mu)=-0.9(2)$ obtained from the negative gluon helicity solution in~\cite{Zhou:2022wzm}. Moreover, the only LQCD calculation~\cite{Yang:2016plb} at the physical pion mass, continuum, and infinite volume limits obtained $\Dl G = 0.251(47)(16)$ using a local matrix element~\cite{Ji:2013fga}. Although the calculation in~\cite{Yang:2016plb} is not free of a large matching systematic error, it is most likely that including various systematics in   future refined calculations will not alter the sign of $\Dl G$. It is remarkable  that $\Delta G$ obtained from this  calculation using nonlocal operator and that obtained using a local operator in~\cite{Yang:2016plb} both result in positive contribution.

Next, we determine $x\Dl g(x,\mu)$  from $\Dl{\mathcal I}_g(\om,\mu)$. Unlike many previous LQCD calculations (for references see~\cite{Constantinou:2022yye}), we avoid constraining the $x$-dependence of PDF using the functional form $x^\alpha(1-x)^\beta$ or an extension to this basic-fit form with one or two  additional parameters. 

\begin{figure}[htp]
	\centering
    \includegraphics[width=3.45in, height=2.1in]{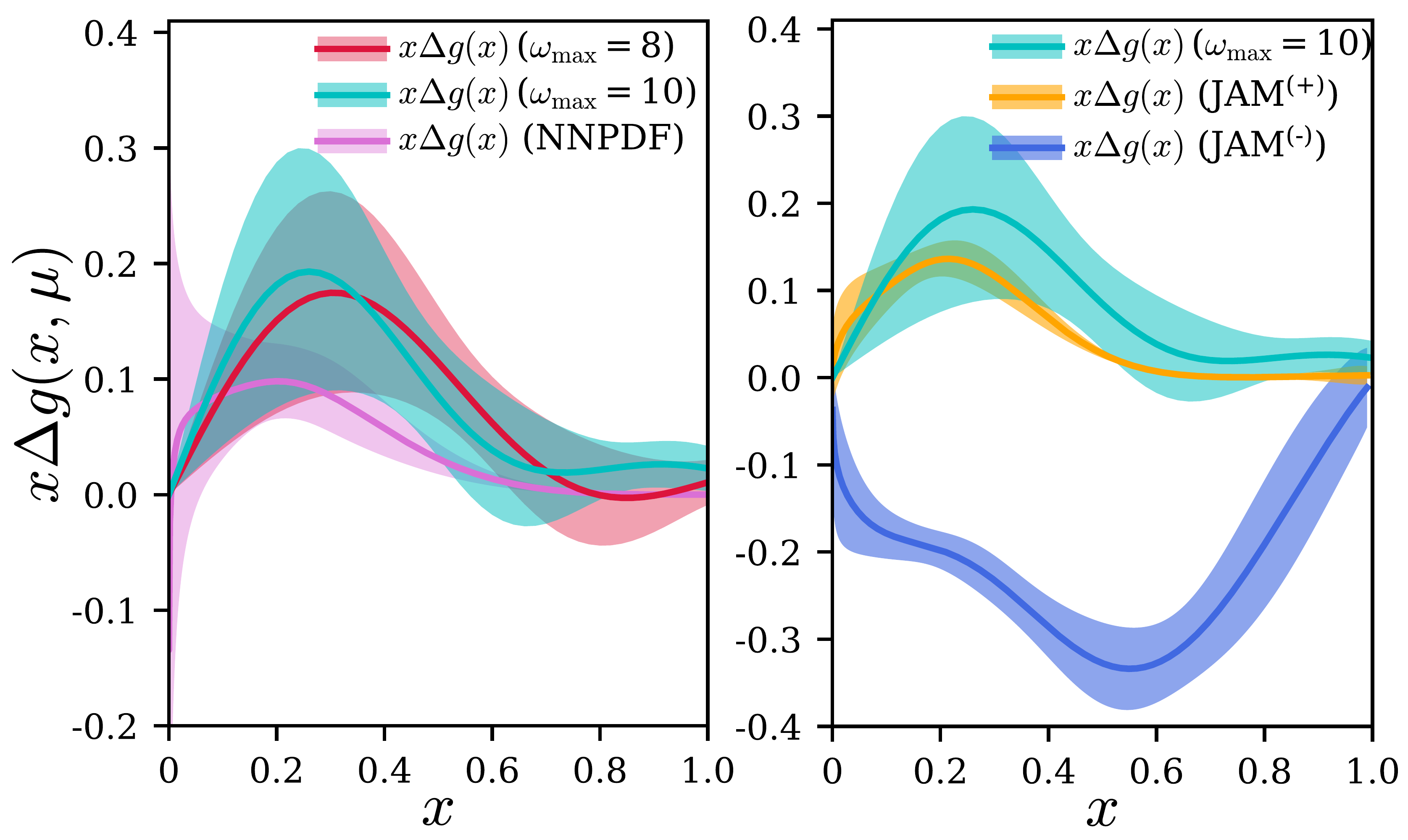}
 	\caption{Functional form independent gluon helicity PDFs (red and cyan bands) from  the lattice data for two different ranges of $\om$ and comparison with the PDFs from NNPDF~\cite{Nocera:2014gqa} (left panel) and  the JAM positive and negative solutions~\cite{Zhou:2022wzm} (right panel) at scale $\mu=2~{\rm GeV}$.} \label{fig:PDF}
 \end{figure}

 It should be emphasized that no functional form  for the PDFs can be successfully used before or without removing the contamination term. Therefore, simple functional forms, such as the $x^\alpha(1-x)^\beta$ ansatz for  the PDFs or its variations, cannot be utilized  to remove the contamination term and  extract  the gluon helicity PDF. This is why  the gluon helicity PDF extraction was not possible in the previous study. Furthermore, as shown above, the moments expansion cannot remove the contamination term over a larger range of Ioffe time, and  the subsequent PDF ansatz fitting cannot extract meaningful PDFs. In addition, for currently available LQCD calculations in  a limited $\om$ range, these functional forms can be biased, leading to unreliable $\chi^2/{\rm d.o.f.}$, and  underestimation of  uncertainties. For example, the same two-parameter form to parametrize $xg(x)$ leads to a diverging PDF  in~\cite{Fan:2020cpa} and a converging PDF in~\cite{HadStruc:2021wmh}, whereas none of these lattice ITDs reach anywhere close to the Regge region or have much sensitivity to the small-$x$ physics. This is true for any LQCD calculation in a limited $\om$ range~\cite{Braun:1994jq,Ji:2013dva}. In~\cite{HadStruc:2021wmh}, $\alpha \geq 0$ constraint was imposed in a Bayesian fit, motivated by a phenomenological analysis in~\cite{Sufian:2020wcv}. Otherwise, it would have resulted in a diverging PDF as in~\cite{Fan:2020cpa}. We, therefore, propose an alternative method to determine $x\Dl g(x)$ from  the lattice data, independent of any functional form of  the PDFs. It can be shown that  $x\Delta g(x)$ is related to $\Delta \mathcal{I}_g(\om,\mu)$ by the following relation:
\bea \label{eq:master3}
x\Delta g(x,\mu) = \frac{2}{\pi} \int_0^{\infty} d\om~ \sin(x\om)~ \Delta \mathcal{I}_g(\om,\mu)\, ,
\eea
which allows us to obtain $x\Dl g(x,\mu)$ at each point in the $x$-space as shown in Fig.~\ref{fig:PDF} without relying on any constraint or prior information. Accuracy of the determination of $x\Delta g(x,\mu)$ in this way depends on the  $\om_{\rm max}$ but gives a true representation of the lattice data  and the extracted PDF exactly reproduces the uncertainty of the ITD. It is assuring to see from Fig.~\ref{fig:matchedITD} that as $\om_{\rm max}$ increases,  $x\Dl g(x,\mu)$ shifts more towards the global analyses results, e.g. the NNPDF and  the JAM$^{(+)}$ fits shown in the figure. With increasing $\om_{\rm max}$, the accuracy of the determination of PDFs can be systematically improved. 

\begin{figure}[htp]
	\centering
 	\includegraphics[width=3.4in, height=2.5in]{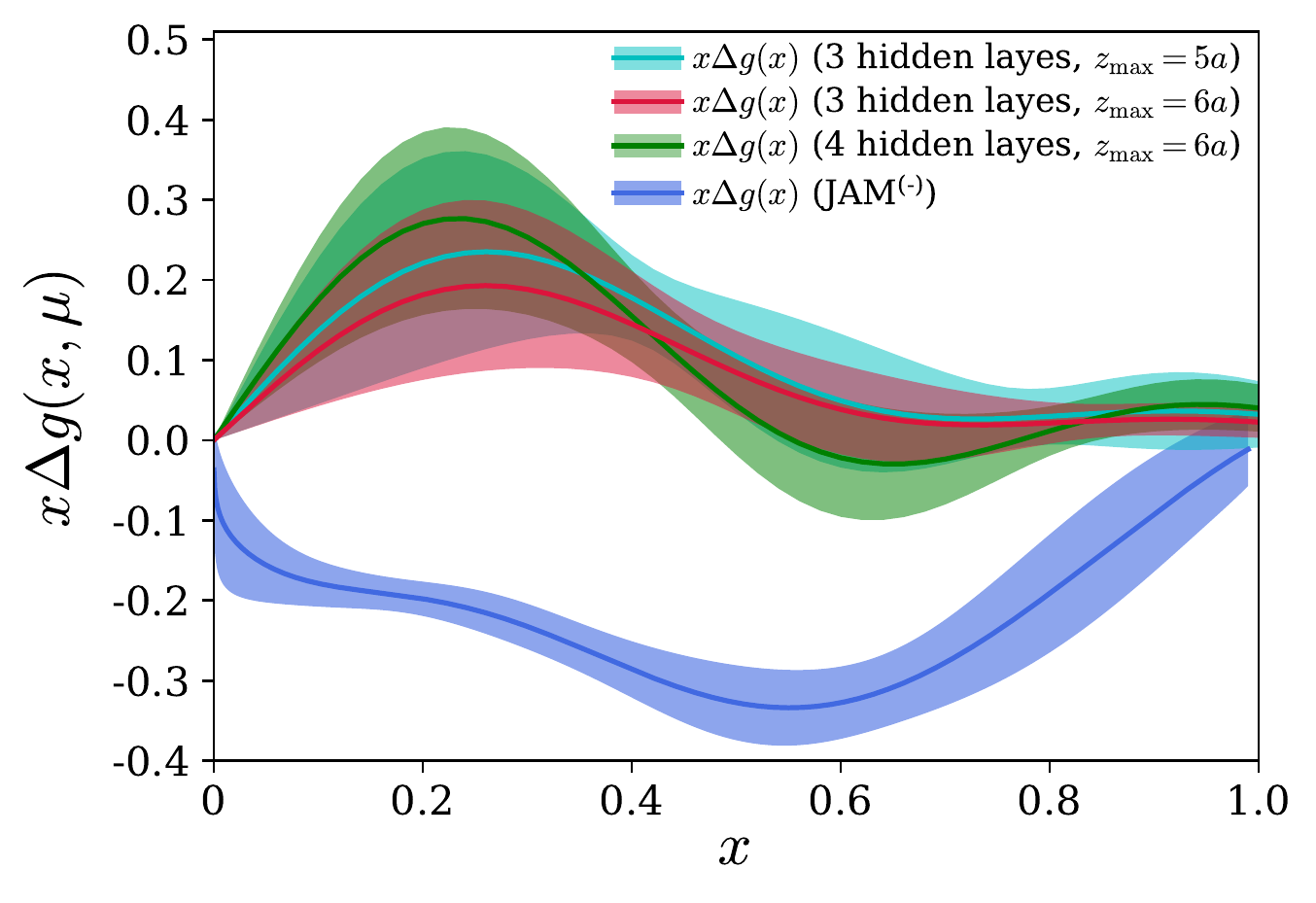}
 	\caption{Variations in the gluon helicity distribution function $x\Delta g(x)$ for varying  the neural network analysis. A comparison is made with the negative gluon helicity solution in Ref.~\cite{Zhou:2022wzm} to demonstrate the preference for a positive gluon helicity solution from this lattice QCD calculation.  }\label{fig:PDFsystematics}
 \end{figure}

 Once again, in Fig.~\ref{fig:PDFsystematics}, we show the gluon helicity PDFs arising from the variations in the neural network analysis discussed above.  The $\Dl G^L(\mu)$ values extracted from the gluon helicity Ioffe time distribution or the gluon helicity PDFs are $0.387( 249)$ (for three hidden layers and $z_{\rm max}=5a$, $0.484( 201)$ (for four hidden layers and $z_{\rm max}=6a$, and $0.405(195)$ (for three hidden layers and $z_{\rm max}=6a$). 
 
\begin{figure}[htp]
	\centering
 	\includegraphics[width=3.4in, height=2.5in]{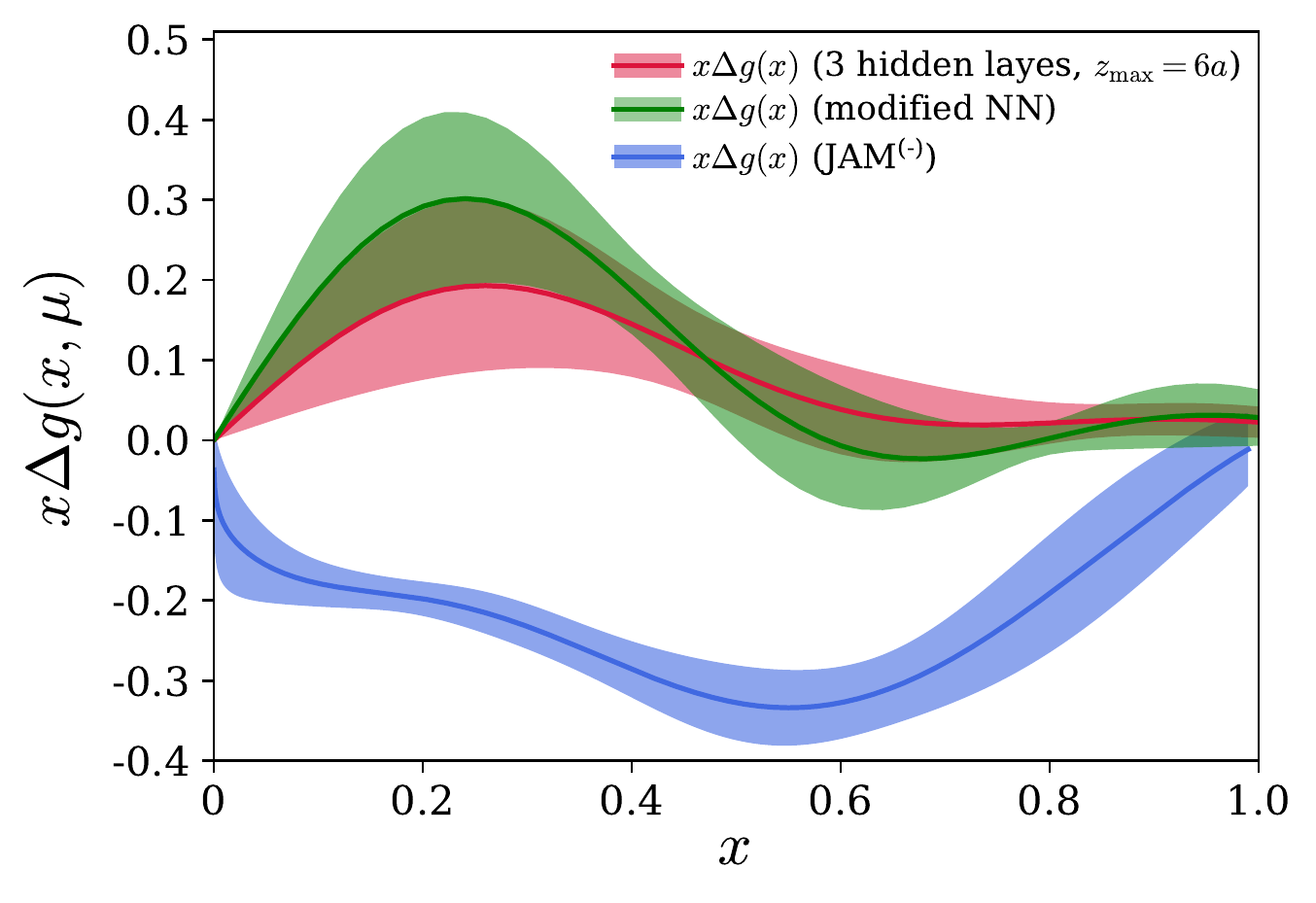}
 	\caption{Variations in the gluon helicity distribution function $x\Delta g(x)$ for varying  the neural network analysis. The NN analysis with four hidden layers and different outputs and activation functions is labeled as the modified neural network  (modified NN) as described in the main text. A comparison is made with the negative gluon helicity solution in Ref.~\cite{Zhou:2022wzm} to demonstrate the preference for a positive gluon helicity solution from this lattice QCD calculation.  }\label{fig:PDFsystematics-alter4}
 \end{figure}

  A similar comparison can be made between the gluon helicity PDFs extracted from the  three-hidden-layer NN and the modified NN with lattice data of $z_{\rm max}=6a$. We see from Fig.~\ref{fig:PDFsystematics-alter4} that in both cases, a positive gluon helicity distribution is preferred from this LQCD calculation, and the behavior of the $x\Dl g(x,\mu)$ distribution does not abruptly change due to the above-mentioned setup of a different neural network analysis. 

It is important to highlight that like any LQCD calculations,  this calculation does not have a solid constraint on the  PDF in the small-$x$  region which is associated with large uncertainties in $\Delta G$ and $x\Dl g(x)$ in global analyses~\cite{Nocera:2014gqa,deFlorian:2014yva} due to  the lack of experimental data. It is the large negative solution found in the moderate to large-$x$ region in~\cite{Zhou:2022wzm}, which is ruled out by our calculation.

\begin{figure}[htp]
	\centering
 	\includegraphics[width=3.4in, height=2.5in]{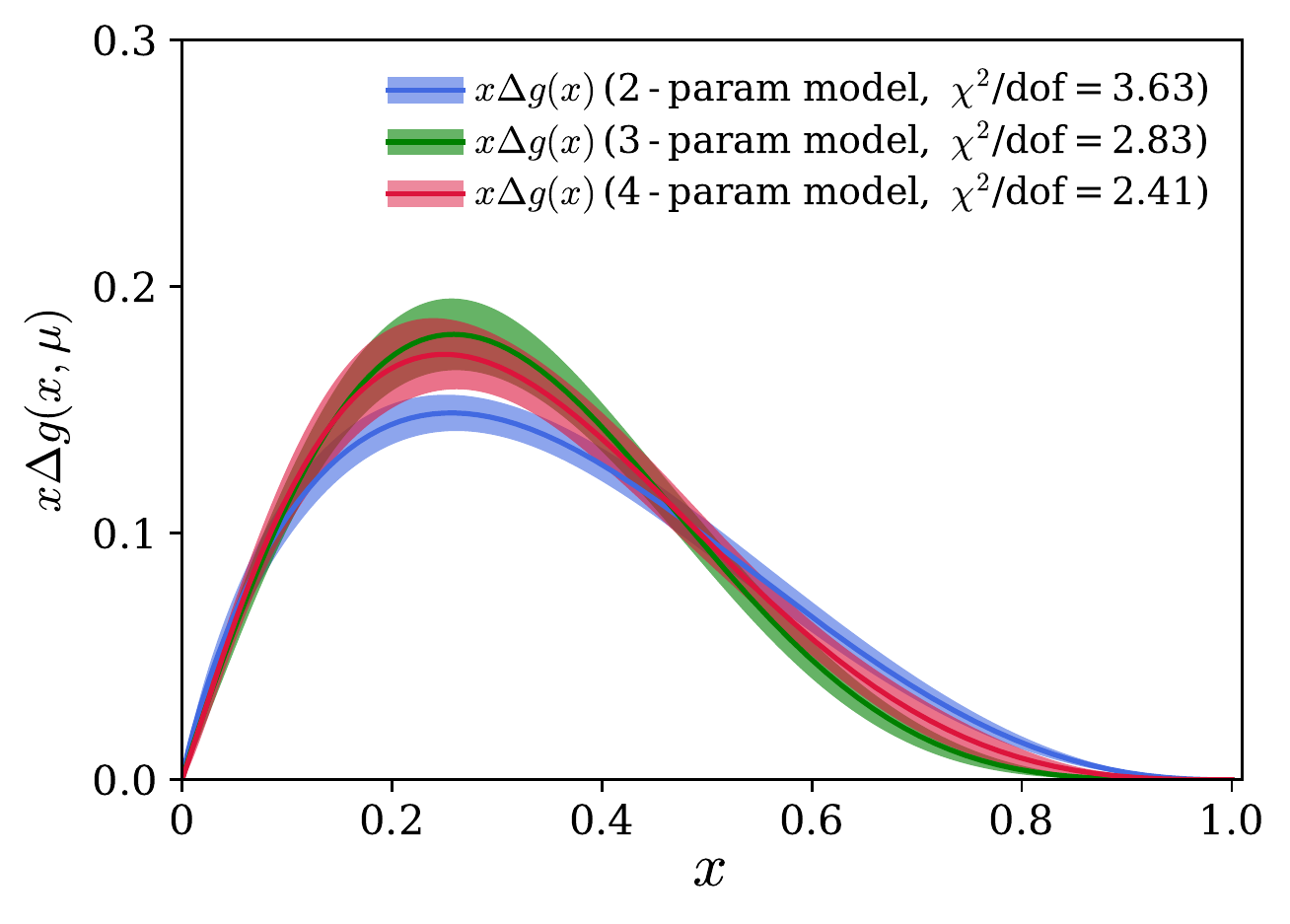}
 	\caption{Model parametrization of $x\Delta g(x,\mu)$ distribution from fit to  $\Delta {\mathcal{I}_g}(\om,\mu)$ data. \label{fig:modelfit}}
 \end{figure}

\begin{figure}[htp]
	\centering
 	\includegraphics[width=3.4in, height=2.5in]{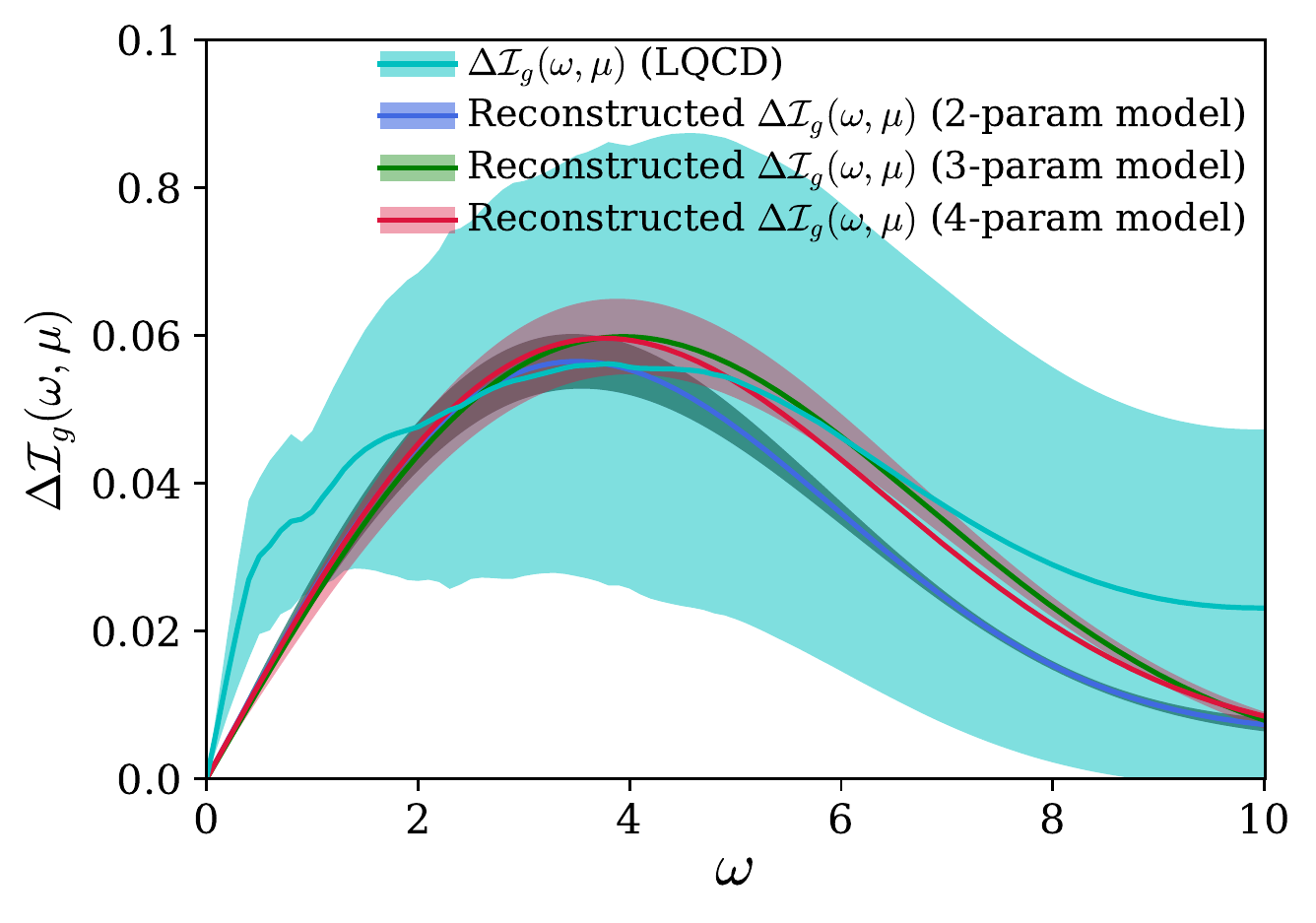}
 	\caption{$\Delta {\mathcal{I}_g}(\om,\mu)$ reconstructed from  the model parametrizations of $x\Delta g(x)$ distributions. The results are compared with  $\Delta {\mathcal{I}_g}(\om,\mu)$ determined from the lattice data (cyan band).\label{fig:modelITD}}
 \end{figure}

 To this end, one can also investigate the determination of $x\Dl g(x)$ by parametrizing  the $\Delta \mathcal{I}_g(\om)$  using a specific functional form, as was done in previous lattice studies~\cite{HadStruc:2021wmh, HadStruc:2022yaw}. We have incorporated the determination of $x\Delta g(x)$ by employing 2, 3, and 4-parameter model fits. For this purpose, we initiate with the 2-parameter fit given by the form $x^\alpha(1-x)^\beta$, and subsequently introduce additional parameters, namely $\rho$ and $\gamma$, resulting in $x^\alpha(1-x)^\beta(1+\rho\sqrt{x}+\gamma x)$.  As anticipated, the derived $x\Delta g(x)$ exhibits significantly reduced uncertainty (accompanied by a larger $\chi^2/{\rm  d.o.f.}$) when compared to its determination through Eq.~\eqref{eq:master3}. This observation has already been discussed in the earlier part of the  paper and can be seen in Refs.~\cite{HadStruc:2021wmh, HadStruc:2022yaw}. For the purpose of an explicit demonstration, we present the results in Fig.~\ref{fig:modelfit}. To provide further illustration, we  show the reconstructed $\Delta {\mathcal{I}_g}(\omega)$ using these constrained model-dependent PDF parametrizations in Fig.~\ref{fig:modelITD}. This motivates us to avoid the model-dependent parametrization of PDFs where uncertainty can be underestimated. We advocate for our approach of deriving the PDF directly from lattice data using Eq.~\eqref{eq:master3}, eliminating the need for additional fitting procedures.

In conclusion,  we have presented the first lattice QCD calculation that supports a positive gluon helicity PDF, and therefore a positive gluon spin contribution to the proton spin budget, in contrast to the most recent global analysis. We  have achieved this by isolating the leading-twist-dominated component $x\Dl g(x,\mu)$ of the gluonic correlation function. This was not possible previously. Instead of relying on the moments expansion or using limited functional forms of  the PDFs on the lattice data, we discuss why a neural network analysis can provide a better solution to extract the gluon helicity PDF. As a future development, this demonstrated interface between LQCD and machine learning can have promising applications to investigate higher-twist contributions at the level of LQCD correlation functions, which is an unexplored research area. We  have also presented a systematically improvable approach to determine  the PDF without relying on any restricted functional forms,  avoiding the possible underestimation of uncertainties of the fitted lattice data and  the PDF. The implementation of these two features will be extended to other quark/gluon correlations in future studies.

 In future investigations, it will be necessary to examine multiple lattice data sets encompassing different lattice spacings, volumes, and pion masses. This will allow for a comprehensive assessment of the diverse systematic uncertainties inherent in this calculation. Currently, the error bands primarily stem from  the statistical and  the neural network analysis uncertainties. Lattice QCD calculations
will continue to improve significantly and with the availability of precise data and calculations conducted on multiple lattice ensembles, the proposed methodology for calculating the gluon helicity PDF  offers essential theoretical support and provides an opportunity to investigate the various systematic uncertainties associated with determining the gluon helicity PDF. Finally, given the current challenges in the extraction of  the gluon helicity from the available experimental data, it is imperative that this lattice QCD calculation can be used to complement global analyses and has the potential to open an avenue for elucidating the role of gluons in the nucleon's spin structure. 
 

\subsection{Acknowledgement}
R.S.S. thanks Nikhil Karthik who provided insight, valuable suggestions, and expertise that greatly assisted this research. R.S.S. acknowledges the mentorship of Robert Edwards, Luka Leskovec, Keh-Fei Liu, Jianwei Qiu, and David Richards and their support and encouragement to pursue this work.  T.L. is supported by the National Natural Science Foundation of China under grant Nos. 12175117 and 20221017-1,  and Shandong Province Natural Science Foundation under grant No. ZFJH202303. R.S.S.  is supported by the U.S. DOE grant No. DE-SC0013065 and DOE grant No. DE-AC05-06OR23177 which is within the framework of the TMD Topical Collaboration. R.S.S. is also  supported by the Special Postdoctoral Researchers Program of RIKEN and RIKEN-BNL Research
Center. The authors also acknowledge partial support by the U.S. Department of Energy, Office of Science, Office of Nuclear Physics under the umbrella of the Quark-Gluon Tomography (QGT) Topical Collaboration with Award No. DE-SC0023646.

\bibliography{References.bib}

\begin{thebibliography}{74}%
\makeatletter
\providecommand \@ifxundefined [1]{%
 \@ifx{#1\undefined}
}%
\providecommand \@ifnum [1]{%
 \ifnum #1\expandafter \@firstoftwo
 \else \expandafter \@secondoftwo
 \fi
}%
\providecommand \@ifx [1]{%
 \ifx #1\expandafter \@firstoftwo
 \else \expandafter \@secondoftwo
 \fi
}%
\providecommand \natexlab [1]{#1}%
\providecommand \enquote  [1]{``#1''}%
\providecommand \bibnamefont  [1]{#1}%
\providecommand \bibfnamefont [1]{#1}%
\providecommand \citenamefont [1]{#1}%
\providecommand \href@noop [0]{\@secondoftwo}%
\providecommand \href [0]{\begingroup \@sanitize@url \@href}%
\providecommand \@href[1]{\@@startlink{#1}\@@href}%
\providecommand \@@href[1]{\endgroup#1\@@endlink}%
\providecommand \@sanitize@url [0]{\catcode `\\12\catcode `\$12\catcode
  `\&12\catcode `\#12\catcode `\^12\catcode `\_12\catcode `\%12\relax}%
\providecommand \@@startlink[1]{}%
\providecommand \@@endlink[0]{}%
\providecommand \url  [0]{\begingroup\@sanitize@url \@url }%
\providecommand \@url [1]{\endgroup\@href {#1}{\urlprefix }}%
\providecommand \urlprefix  [0]{URL }%
\providecommand \Eprint [0]{\href }%
\providecommand \doibase [0]{http://dx.doi.org/}%
\providecommand \selectlanguage [0]{\@gobble}%
\providecommand \bibinfo  [0]{\@secondoftwo}%
\providecommand \bibfield  [0]{\@secondoftwo}%
\providecommand \translation [1]{[#1]}%
\providecommand \BibitemOpen [0]{}%
\providecommand \bibitemStop [0]{}%
\providecommand \bibitemNoStop [0]{.\EOS\space}%
\providecommand \EOS [0]{\spacefactor3000\relax}%
\providecommand \BibitemShut  [1]{\csname bibitem#1\endcsname}%
\let\auto@bib@innerbib\@empty
\bibitem [{\citenamefont {Ashman}\ \emph {et~al.}(1988)\citenamefont {Ashman}
  \emph {et~al.}}]{EuropeanMuon:1987isl}%
  \BibitemOpen
  \bibfield  {author} {\bibinfo {author} {\bibfnamefont {J.}~\bibnamefont
  {Ashman}} \emph {et~al.} (\bibinfo {collaboration} {European Muon
  Collaboration}),\ }\href {\doibase 10.1016/0370-2693(88)91523-7} {\bibfield
  {journal} {\bibinfo  {journal} {Phys. Lett. B}\ }\textbf {\bibinfo {volume}
  {206}},\ \bibinfo {pages} {364} (\bibinfo {year} {1988})}\BibitemShut
  {NoStop}%
\bibitem [{\citenamefont {Ashman}\ \emph {et~al.}(1989)\citenamefont {Ashman}
  \emph {et~al.}}]{EuropeanMuon:1989yki}%
  \BibitemOpen
  \bibfield  {author} {\bibinfo {author} {\bibfnamefont {J.}~\bibnamefont
  {Ashman}} \emph {et~al.} (\bibinfo {collaboration} {European Muon
  Collaboration}),\ }\href {\doibase 10.1016/0550-3213(89)90089-8} {\bibfield
  {journal} {\bibinfo  {journal} {Nucl. Phys. B}\ }\textbf {\bibinfo {volume}
  {328}},\ \bibinfo {pages} {1} (\bibinfo {year} {1989})}\BibitemShut {NoStop}%
\bibitem [{\citenamefont {Bunce}\ \emph {et~al.}(2000)\citenamefont {Bunce},
  \citenamefont {Saito}, \citenamefont {Soffer},\ and\ \citenamefont
  {Vogelsang}}]{Bunce:2000uv}%
  \BibitemOpen
  \bibfield  {author} {\bibinfo {author} {\bibfnamefont {G.}~\bibnamefont
  {Bunce}}, \bibinfo {author} {\bibfnamefont {N.}~\bibnamefont {Saito}},
  \bibinfo {author} {\bibfnamefont {J.}~\bibnamefont {Soffer}}, \ and\ \bibinfo
  {author} {\bibfnamefont {W.}~\bibnamefont {Vogelsang}},\ }\href {\doibase
  10.1146/annurev.nucl.50.1.525} {\bibfield  {journal} {\bibinfo  {journal}
  {Ann. Rev. Nucl. Part. Sci.}\ }\textbf {\bibinfo {volume} {50}},\ \bibinfo
  {pages} {525} (\bibinfo {year} {2000})},\ \Eprint
  {http://arxiv.org/abs/hep-ph/0007218} {arXiv:hep-ph/0007218} \BibitemShut
  {NoStop}%
\bibitem [{\citenamefont {Adare}\ \emph {et~al.}(2014)\citenamefont {Adare}
  \emph {et~al.}}]{PHENIX:2014gbf}%
  \BibitemOpen
  \bibfield  {author} {\bibinfo {author} {\bibfnamefont {A.}~\bibnamefont
  {Adare}} \emph {et~al.} (\bibinfo {collaboration} {PHENIX Collaboration}),\
  }\href {\doibase 10.1103/PhysRevD.90.012007} {\bibfield  {journal} {\bibinfo
  {journal} {Phys. Rev. D}\ }\textbf {\bibinfo {volume} {90}},\ \bibinfo
  {pages} {012007} (\bibinfo {year} {2014})},\ \Eprint
  {http://arxiv.org/abs/1402.6296} {arXiv:1402.6296 [hep-ex]} \BibitemShut
  {NoStop}%
\bibitem [{\citenamefont {Airapetian}\ \emph {et~al.}(2008)\citenamefont
  {Airapetian} \emph {et~al.}}]{HERMES:2008abz}%
  \BibitemOpen
  \bibfield  {author} {\bibinfo {author} {\bibfnamefont {A.}~\bibnamefont
  {Airapetian}} \emph {et~al.} (\bibinfo {collaboration} {HERMES
  Collaboration}),\ }\href {\doibase 10.1088/1126-6708/2008/06/066} {\bibfield
  {journal} {\bibinfo  {journal} {JHEP}\ }\textbf {\bibinfo {volume} {06}},\
  \bibinfo {pages} {066} (\bibinfo {year} {2008})},\ \Eprint
  {http://arxiv.org/abs/0802.2499} {arXiv:0802.2499 [hep-ex]} \BibitemShut
  {NoStop}%
\bibitem [{\citenamefont {Akhunzyanov}\ \emph {et~al.}(2019)\citenamefont
  {Akhunzyanov} \emph {et~al.}}]{COMPASS:2018pup}%
  \BibitemOpen
  \bibfield  {author} {\bibinfo {author} {\bibfnamefont {R.}~\bibnamefont
  {Akhunzyanov}} \emph {et~al.} (\bibinfo {collaboration} {COMPASS
  Collaboration}),\ }\href {\doibase 10.1016/j.physletb.2019.04.038} {\bibfield
   {journal} {\bibinfo  {journal} {Phys. Lett. B}\ }\textbf {\bibinfo {volume}
  {793}},\ \bibinfo {pages} {188} (\bibinfo {year} {2019})},\ \bibinfo {note}
  {[Erratum: \href{https://doi.org/10.1016/j.physletb.2019.135129}{Phys. Lett.
  B {\bf 800}, 135129 (2020)}]},\ \Eprint {http://arxiv.org/abs/1802.02739}
  {arXiv:1802.02739 [hep-ex]} \BibitemShut {NoStop}%
\bibitem [{\citenamefont {Accardi}\ \emph {et~al.}(2016)\citenamefont {Accardi}
  \emph {et~al.}}]{Accardi:2012qut}%
  \BibitemOpen
  \bibfield  {author} {\bibinfo {author} {\bibfnamefont {A.}~\bibnamefont
  {Accardi}} \emph {et~al.},\ }\href {\doibase 10.1140/epja/i2016-16268-9}
  {\bibfield  {journal} {\bibinfo  {journal} {Eur. Phys. J. A}\ }\textbf
  {\bibinfo {volume} {52}},\ \bibinfo {pages} {268} (\bibinfo {year} {2016})},\
  \Eprint {http://arxiv.org/abs/1212.1701} {arXiv:1212.1701 [nucl-ex]}
  \BibitemShut {NoStop}%
\bibitem [{\citenamefont {Abdul~Khalek}\ \emph {et~al.}(2022)\citenamefont
  {Abdul~Khalek} \emph {et~al.}}]{AbdulKhalek:2021gbh}%
  \BibitemOpen
  \bibfield  {author} {\bibinfo {author} {\bibfnamefont {R.}~\bibnamefont
  {Abdul~Khalek}} \emph {et~al.},\ }\href {\doibase
  10.1016/j.nuclphysa.2022.122447} {\bibfield  {journal} {\bibinfo  {journal}
  {Nucl. Phys. A}\ }\textbf {\bibinfo {volume} {1026}},\ \bibinfo {pages}
  {122447} (\bibinfo {year} {2022})},\ \Eprint
  {http://arxiv.org/abs/2103.05419} {arXiv:2103.05419 [physics.ins-det]}
  \BibitemShut {NoStop}%
\bibitem [{\citenamefont {Anderle}\ \emph {et~al.}(2021)\citenamefont {Anderle}
  \emph {et~al.}}]{Anderle:2021wcy}%
  \BibitemOpen
  \bibfield  {author} {\bibinfo {author} {\bibfnamefont {D.~P.}\ \bibnamefont
  {Anderle}} \emph {et~al.},\ }\href {\doibase 10.1007/s11467-021-1062-0}
  {\bibfield  {journal} {\bibinfo  {journal} {Front. Phys. (Beijing)}\ }\textbf
  {\bibinfo {volume} {16}},\ \bibinfo {pages} {64701} (\bibinfo {year}
  {2021})},\ \Eprint {http://arxiv.org/abs/2102.09222} {arXiv:2102.09222
  [nucl-ex]} \BibitemShut {NoStop}%
\bibitem [{\citenamefont {Jaffe}\ and\ \citenamefont
  {Manohar}(1990)}]{Jaffe:1989jz}%
  \BibitemOpen
  \bibfield  {author} {\bibinfo {author} {\bibfnamefont {R.~L.}\ \bibnamefont
  {Jaffe}}\ and\ \bibinfo {author} {\bibfnamefont {A.}~\bibnamefont
  {Manohar}},\ }\href {\doibase 10.1016/0550-3213(90)90506-9} {\bibfield
  {journal} {\bibinfo  {journal} {Nucl. Phys. B}\ }\textbf {\bibinfo {volume}
  {337}},\ \bibinfo {pages} {509} (\bibinfo {year} {1990})}\BibitemShut
  {NoStop}%
\bibitem [{\citenamefont {Ji}(1997)}]{Ji:1996ek}%
  \BibitemOpen
  \bibfield  {author} {\bibinfo {author} {\bibfnamefont {X.-D.}\ \bibnamefont
  {Ji}},\ }\href {\doibase 10.1103/PhysRevLett.78.610} {\bibfield  {journal}
  {\bibinfo  {journal} {Phys. Rev. Lett.}\ }\textbf {\bibinfo {volume} {78}},\
  \bibinfo {pages} {610} (\bibinfo {year} {1997})},\ \Eprint
  {http://arxiv.org/abs/hep-ph/9603249} {arXiv:hep-ph/9603249} \BibitemShut
  {NoStop}%
\bibitem [{\citenamefont {Aidala}\ \emph {et~al.}(2013)\citenamefont {Aidala},
  \citenamefont {Bass}, \citenamefont {Hasch},\ and\ \citenamefont
  {Mallot}}]{Aidala:2012mv}%
  \BibitemOpen
  \bibfield  {author} {\bibinfo {author} {\bibfnamefont {C.~A.}\ \bibnamefont
  {Aidala}}, \bibinfo {author} {\bibfnamefont {S.~D.}\ \bibnamefont {Bass}},
  \bibinfo {author} {\bibfnamefont {D.}~\bibnamefont {Hasch}}, \ and\ \bibinfo
  {author} {\bibfnamefont {G.~K.}\ \bibnamefont {Mallot}},\ }\href {\doibase
  10.1103/RevModPhys.85.655} {\bibfield  {journal} {\bibinfo  {journal} {Rev.
  Mod. Phys.}\ }\textbf {\bibinfo {volume} {85}},\ \bibinfo {pages} {655}
  (\bibinfo {year} {2013})},\ \Eprint {http://arxiv.org/abs/1209.2803}
  {arXiv:1209.2803 [hep-ph]} \BibitemShut {NoStop}%
\bibitem [{\citenamefont {Leader}\ and\ \citenamefont
  {Lorc\'e}(2014)}]{Leader:2013jra}%
  \BibitemOpen
  \bibfield  {author} {\bibinfo {author} {\bibfnamefont {E.}~\bibnamefont
  {Leader}}\ and\ \bibinfo {author} {\bibfnamefont {C.}~\bibnamefont
  {Lorc\'e}},\ }\href {\doibase 10.1016/j.physrep.2014.02.010} {\bibfield
  {journal} {\bibinfo  {journal} {Phys. Rep.}\ }\textbf {\bibinfo {volume}
  {541}},\ \bibinfo {pages} {163} (\bibinfo {year} {2014})},\ \Eprint
  {http://arxiv.org/abs/1309.4235} {arXiv:1309.4235 [hep-ph]} \BibitemShut
  {NoStop}%
\bibitem [{\citenamefont {Wakamatsu}(2014)}]{Wakamatsu:2014zza}%
  \BibitemOpen
  \bibfield  {author} {\bibinfo {author} {\bibfnamefont {M.}~\bibnamefont
  {Wakamatsu}},\ }\href {\doibase 10.1142/S0217751X14300129} {\bibfield
  {journal} {\bibinfo  {journal} {Int. J. Mod. Phys. A}\ }\textbf {\bibinfo
  {volume} {29}},\ \bibinfo {pages} {1430012} (\bibinfo {year} {2014})},\
  \Eprint {http://arxiv.org/abs/1402.4193} {arXiv:1402.4193 [hep-ph]}
  \BibitemShut {NoStop}%
\bibitem [{\citenamefont {Deur}\ \emph {et~al.}(2018)\citenamefont {Deur},
  \citenamefont {Brodsky},\ and\ \citenamefont {de~T\'eramond}}]{Deur:2018roz}%
  \BibitemOpen
  \bibfield  {author} {\bibinfo {author} {\bibfnamefont {A.}~\bibnamefont
  {Deur}}, \bibinfo {author} {\bibfnamefont {S.~J.}\ \bibnamefont {Brodsky}}, \
  and\ \bibinfo {author} {\bibfnamefont {G.~F.}\ \bibnamefont
  {de~T\'eramond}},\ }\href {\doibase 10.1088/1361-6633/ab0b8f} {\bibfield
  {journal} {\bibinfo  {journal} {Rep. Prog. Phys.}\ }\textbf {\bibinfo
  {volume} {82}},\ \bibinfo {pages} {076201} (\bibinfo {year} {2018})},\
  \Eprint {http://arxiv.org/abs/1807.05250} {arXiv:1807.05250 [hep-ph]}
  \BibitemShut {NoStop}%
\bibitem [{\citenamefont {Ji}\ \emph {et~al.}(2021{\natexlab{a}})\citenamefont
  {Ji}, \citenamefont {Yuan},\ and\ \citenamefont {Zhao}}]{Ji:2020ena}%
  \BibitemOpen
  \bibfield  {author} {\bibinfo {author} {\bibfnamefont {X.}~\bibnamefont
  {Ji}}, \bibinfo {author} {\bibfnamefont {F.}~\bibnamefont {Yuan}}, \ and\
  \bibinfo {author} {\bibfnamefont {Y.}~\bibnamefont {Zhao}},\ }\href {\doibase
  10.1038/s42254-020-00248-4} {\bibfield  {journal} {\bibinfo  {journal}
  {Nature Rev. Phys.}\ }\textbf {\bibinfo {volume} {3}},\ \bibinfo {pages} {27}
  (\bibinfo {year} {2021}{\natexlab{a}})},\ \Eprint
  {http://arxiv.org/abs/2009.01291} {arXiv:2009.01291 [hep-ph]} \BibitemShut
  {NoStop}%
\bibitem [{\citenamefont {Liu}(2022)}]{Liu:2021lke}%
  \BibitemOpen
  \bibfield  {author} {\bibinfo {author} {\bibfnamefont {K.-F.}\ \bibnamefont
  {Liu}},\ }\href {\doibase 10.1007/s43673-022-00037-4} {\bibfield  {journal}
  {\bibinfo  {journal} {AAPPS Bull.}\ }\textbf {\bibinfo {volume} {32}},\
  \bibinfo {pages} {8} (\bibinfo {year} {2022})},\ \Eprint
  {http://arxiv.org/abs/2112.08416} {arXiv:2112.08416 [hep-lat]} \BibitemShut
  {NoStop}%
\bibitem [{\citenamefont {de~Florian}\ \emph {et~al.}(2009)\citenamefont
  {de~Florian}, \citenamefont {Sassot}, \citenamefont {Stratmann},\ and\
  \citenamefont {Vogelsang}}]{deFlorian:2009vb}%
  \BibitemOpen
  \bibfield  {author} {\bibinfo {author} {\bibfnamefont {D.}~\bibnamefont
  {de~Florian}}, \bibinfo {author} {\bibfnamefont {R.}~\bibnamefont {Sassot}},
  \bibinfo {author} {\bibfnamefont {M.}~\bibnamefont {Stratmann}}, \ and\
  \bibinfo {author} {\bibfnamefont {W.}~\bibnamefont {Vogelsang}},\ }\href
  {\doibase 10.1103/PhysRevD.80.034030} {\bibfield  {journal} {\bibinfo
  {journal} {Phys. Rev. D}\ }\textbf {\bibinfo {volume} {80}},\ \bibinfo
  {pages} {034030} (\bibinfo {year} {2009})},\ \Eprint
  {http://arxiv.org/abs/0904.3821} {arXiv:0904.3821 [hep-ph]} \BibitemShut
  {NoStop}%
\bibitem [{\citenamefont {Nocera}\ \emph {et~al.}(2014)\citenamefont {Nocera},
  \citenamefont {Ball}, \citenamefont {Forte}, \citenamefont {Ridolfi},\ and\
  \citenamefont {Rojo}}]{Nocera:2014gqa}%
  \BibitemOpen
  \bibfield  {author} {\bibinfo {author} {\bibfnamefont {E.~R.}\ \bibnamefont
  {Nocera}}, \bibinfo {author} {\bibfnamefont {R.~D.}\ \bibnamefont {Ball}},
  \bibinfo {author} {\bibfnamefont {S.}~\bibnamefont {Forte}}, \bibinfo
  {author} {\bibfnamefont {G.}~\bibnamefont {Ridolfi}}, \ and\ \bibinfo
  {author} {\bibfnamefont {J.}~\bibnamefont {Rojo}} (\bibinfo {collaboration}
  {NNPDF Collaboration}),\ }\href {\doibase 10.1016/j.nuclphysb.2014.08.008}
  {\bibfield  {journal} {\bibinfo  {journal} {Nucl. Phys. B}\ }\textbf
  {\bibinfo {volume} {887}},\ \bibinfo {pages} {276} (\bibinfo {year}
  {2014})},\ \Eprint {http://arxiv.org/abs/1406.5539} {arXiv:1406.5539
  [hep-ph]} \BibitemShut {NoStop}%
\bibitem [{\citenamefont {Adolph}\ \emph {et~al.}(2016)\citenamefont {Adolph}
  \emph {et~al.}}]{COMPASS:2015mhb}%
  \BibitemOpen
  \bibfield  {author} {\bibinfo {author} {\bibfnamefont {C.}~\bibnamefont
  {Adolph}} \emph {et~al.} (\bibinfo {collaboration} {COMPASS Collaboration}),\
  }\href {\doibase 10.1016/j.physletb.2015.11.064} {\bibfield  {journal}
  {\bibinfo  {journal} {Phys. Lett. B}\ }\textbf {\bibinfo {volume} {753}},\
  \bibinfo {pages} {18} (\bibinfo {year} {2016})},\ \Eprint
  {http://arxiv.org/abs/1503.08935} {arXiv:1503.08935 [hep-ex]} \BibitemShut
  {NoStop}%
\bibitem [{\citenamefont {Ethier}\ \emph {et~al.}(2017)\citenamefont {Ethier},
  \citenamefont {Sato},\ and\ \citenamefont {Melnitchouk}}]{Ethier:2017zbq}%
  \BibitemOpen
  \bibfield  {author} {\bibinfo {author} {\bibfnamefont {J.~J.}\ \bibnamefont
  {Ethier}}, \bibinfo {author} {\bibfnamefont {N.}~\bibnamefont {Sato}}, \ and\
  \bibinfo {author} {\bibfnamefont {W.}~\bibnamefont {Melnitchouk}},\ }\href
  {\doibase 10.1103/PhysRevLett.119.132001} {\bibfield  {journal} {\bibinfo
  {journal} {Phys. Rev. Lett.}\ }\textbf {\bibinfo {volume} {119}},\ \bibinfo
  {pages} {132001} (\bibinfo {year} {2017})},\ \Eprint
  {http://arxiv.org/abs/1705.05889} {arXiv:1705.05889 [hep-ph]} \BibitemShut
  {NoStop}%
\bibitem [{\citenamefont {Lin}\ \emph {et~al.}(2018)\citenamefont {Lin},
  \citenamefont {Gupta}, \citenamefont {Yoon}, \citenamefont {Jang},\ and\
  \citenamefont {Bhattacharya}}]{Lin:2018obj}%
  \BibitemOpen
  \bibfield  {author} {\bibinfo {author} {\bibfnamefont {H.-W.}\ \bibnamefont
  {Lin}}, \bibinfo {author} {\bibfnamefont {R.}~\bibnamefont {Gupta}}, \bibinfo
  {author} {\bibfnamefont {B.}~\bibnamefont {Yoon}}, \bibinfo {author}
  {\bibfnamefont {Y.-C.}\ \bibnamefont {Jang}}, \ and\ \bibinfo {author}
  {\bibfnamefont {T.}~\bibnamefont {Bhattacharya}},\ }\href {\doibase
  10.1103/PhysRevD.98.094512} {\bibfield  {journal} {\bibinfo  {journal} {Phys.
  Rev. D}\ }\textbf {\bibinfo {volume} {98}},\ \bibinfo {pages} {094512}
  (\bibinfo {year} {2018})},\ \Eprint {http://arxiv.org/abs/1806.10604}
  {arXiv:1806.10604 [hep-lat]} \BibitemShut {NoStop}%
\bibitem [{\citenamefont {Alexandrou}\ \emph {et~al.}(2020)\citenamefont
  {Alexandrou}, \citenamefont {Bacchio}, \citenamefont {Constantinou},
  \citenamefont {Finkenrath}, \citenamefont {Hadjiyiannakou}, \citenamefont
  {Jansen}, \citenamefont {Koutsou}, \citenamefont {Panagopoulos},\ and\
  \citenamefont {Spanoudes}}]{Alexandrou:2020sml}%
  \BibitemOpen
  \bibfield  {author} {\bibinfo {author} {\bibfnamefont {C.}~\bibnamefont
  {Alexandrou}}, \bibinfo {author} {\bibfnamefont {S.}~\bibnamefont {Bacchio}},
  \bibinfo {author} {\bibfnamefont {M.}~\bibnamefont {Constantinou}}, \bibinfo
  {author} {\bibfnamefont {J.}~\bibnamefont {Finkenrath}}, \bibinfo {author}
  {\bibfnamefont {K.}~\bibnamefont {Hadjiyiannakou}}, \bibinfo {author}
  {\bibfnamefont {K.}~\bibnamefont {Jansen}}, \bibinfo {author} {\bibfnamefont
  {G.}~\bibnamefont {Koutsou}}, \bibinfo {author} {\bibfnamefont
  {H.}~\bibnamefont {Panagopoulos}}, \ and\ \bibinfo {author} {\bibfnamefont
  {G.}~\bibnamefont {Spanoudes}},\ }\href {\doibase
  10.1103/PhysRevD.101.094513} {\bibfield  {journal} {\bibinfo  {journal}
  {Phys. Rev. D}\ }\textbf {\bibinfo {volume} {101}},\ \bibinfo {pages}
  {094513} (\bibinfo {year} {2020})},\ \Eprint
  {http://arxiv.org/abs/2003.08486} {arXiv:2003.08486 [hep-lat]} \BibitemShut
  {NoStop}%
\bibitem [{\citenamefont {Wang}\ \emph {et~al.}(2022)\citenamefont {Wang},
  \citenamefont {Yang}, \citenamefont {Liang}, \citenamefont {Draper},\ and\
  \citenamefont {Liu}}]{Wang:2021vqy}%
  \BibitemOpen
  \bibfield  {author} {\bibinfo {author} {\bibfnamefont {G.}~\bibnamefont
  {Wang}}, \bibinfo {author} {\bibfnamefont {Y.-B.}\ \bibnamefont {Yang}},
  \bibinfo {author} {\bibfnamefont {J.}~\bibnamefont {Liang}}, \bibinfo
  {author} {\bibfnamefont {T.}~\bibnamefont {Draper}}, \ and\ \bibinfo {author}
  {\bibfnamefont {K.-F.}\ \bibnamefont {Liu}} (\bibinfo {collaboration}
  {\ensuremath{\chi}QCD}),\ }\href {\doibase 10.1103/PhysRevD.106.014512}
  {\bibfield  {journal} {\bibinfo  {journal} {Phys. Rev. D}\ }\textbf {\bibinfo
  {volume} {106}},\ \bibinfo {pages} {014512} (\bibinfo {year} {2022})},\
  \Eprint {http://arxiv.org/abs/2111.09329} {arXiv:2111.09329 [hep-lat]}
  \BibitemShut {NoStop}%
\bibitem [{\citenamefont {de~Florian}\ \emph {et~al.}(2014)\citenamefont
  {de~Florian}, \citenamefont {Sassot}, \citenamefont {Stratmann},\ and\
  \citenamefont {Vogelsang}}]{deFlorian:2014yva}%
  \BibitemOpen
  \bibfield  {author} {\bibinfo {author} {\bibfnamefont {D.}~\bibnamefont
  {de~Florian}}, \bibinfo {author} {\bibfnamefont {R.}~\bibnamefont {Sassot}},
  \bibinfo {author} {\bibfnamefont {M.}~\bibnamefont {Stratmann}}, \ and\
  \bibinfo {author} {\bibfnamefont {W.}~\bibnamefont {Vogelsang}},\ }\href
  {\doibase 10.1103/PhysRevLett.113.012001} {\bibfield  {journal} {\bibinfo
  {journal} {Phys. Rev. Lett.}\ }\textbf {\bibinfo {volume} {113}},\ \bibinfo
  {pages} {012001} (\bibinfo {year} {2014})},\ \Eprint
  {http://arxiv.org/abs/1404.4293} {arXiv:1404.4293 [hep-ph]} \BibitemShut
  {NoStop}%
\bibitem [{\citenamefont {Zhou}\ \emph {et~al.}(2022)\citenamefont {Zhou},
  \citenamefont {Sato},\ and\ \citenamefont {Melnitchouk}}]{Zhou:2022wzm}%
  \BibitemOpen
  \bibfield  {author} {\bibinfo {author} {\bibfnamefont {Y.}~\bibnamefont
  {Zhou}}, \bibinfo {author} {\bibfnamefont {N.}~\bibnamefont {Sato}}, \ and\
  \bibinfo {author} {\bibfnamefont {W.}~\bibnamefont {Melnitchouk}},\ }\href
  {\doibase 10.1103/PhysRevD.105.074022} {\bibfield  {journal} {\bibinfo
  {journal} {Phys. Rev. D}\ }\textbf {\bibinfo {volume} {105}},\ \bibinfo
  {pages} {074022} (\bibinfo {year} {2022})},\ \Eprint
  {http://arxiv.org/abs/2201.02075} {arXiv:2201.02075 [hep-ph]} \BibitemShut
  {NoStop}%
\bibitem [{\citenamefont {Whitehill}\ \emph {et~al.}(2022)\citenamefont
  {Whitehill}, \citenamefont {Zhou}, \citenamefont {Sato},\ and\ \citenamefont
  {Melnitchouk}}]{Whitehill:2022mpq}%
  \BibitemOpen
  \bibfield  {author} {\bibinfo {author} {\bibfnamefont {R.~M.}\ \bibnamefont
  {Whitehill}}, \bibinfo {author} {\bibfnamefont {Y.}~\bibnamefont {Zhou}},
  \bibinfo {author} {\bibfnamefont {N.}~\bibnamefont {Sato}}, \ and\ \bibinfo
  {author} {\bibfnamefont {W.}~\bibnamefont {Melnitchouk}},\ }\href@noop {} {\
  (\bibinfo {year} {2022})},\ \Eprint {http://arxiv.org/abs/2210.12295}
  {arXiv:2210.12295 [hep-ph]} \BibitemShut {NoStop}%
\bibitem [{\citenamefont {Aschenauer}\ \emph {et~al.}(2015)\citenamefont
  {Aschenauer}, \citenamefont {Sassot},\ and\ \citenamefont
  {Stratmann}}]{Aschenauer:2015ata}%
  \BibitemOpen
  \bibfield  {author} {\bibinfo {author} {\bibfnamefont {E.~C.}\ \bibnamefont
  {Aschenauer}}, \bibinfo {author} {\bibfnamefont {R.}~\bibnamefont {Sassot}},
  \ and\ \bibinfo {author} {\bibfnamefont {M.}~\bibnamefont {Stratmann}},\
  }\href {\doibase 10.1103/PhysRevD.92.094030} {\bibfield  {journal} {\bibinfo
  {journal} {Phys. Rev. D}\ }\textbf {\bibinfo {volume} {92}},\ \bibinfo
  {pages} {094030} (\bibinfo {year} {2015})},\ \Eprint
  {http://arxiv.org/abs/1509.06489} {arXiv:1509.06489 [hep-ph]} \BibitemShut
  {NoStop}%
\bibitem [{\citenamefont {Borsa}\ \emph {et~al.}(2020)\citenamefont {Borsa},
  \citenamefont {Lucero}, \citenamefont {Sassot}, \citenamefont {Aschenauer},\
  and\ \citenamefont {Nunes}}]{Borsa:2020lsz}%
  \BibitemOpen
  \bibfield  {author} {\bibinfo {author} {\bibfnamefont {I.}~\bibnamefont
  {Borsa}}, \bibinfo {author} {\bibfnamefont {G.}~\bibnamefont {Lucero}},
  \bibinfo {author} {\bibfnamefont {R.}~\bibnamefont {Sassot}}, \bibinfo
  {author} {\bibfnamefont {E.~C.}\ \bibnamefont {Aschenauer}}, \ and\ \bibinfo
  {author} {\bibfnamefont {A.~S.}\ \bibnamefont {Nunes}},\ }\href {\doibase
  10.1103/PhysRevD.102.094018} {\bibfield  {journal} {\bibinfo  {journal}
  {Phys. Rev. D}\ }\textbf {\bibinfo {volume} {102}},\ \bibinfo {pages}
  {094018} (\bibinfo {year} {2020})},\ \Eprint
  {http://arxiv.org/abs/2007.08300} {arXiv:2007.08300 [hep-ph]} \BibitemShut
  {NoStop}%
\bibitem [{\citenamefont {Liu}\ and\ \citenamefont {Dong}(1994)}]{Liu:1993cv}%
  \BibitemOpen
  \bibfield  {author} {\bibinfo {author} {\bibfnamefont {K.-F.}\ \bibnamefont
  {Liu}}\ and\ \bibinfo {author} {\bibfnamefont {S.-J.}\ \bibnamefont {Dong}},\
  }\href {\doibase 10.1103/PhysRevLett.72.1790} {\bibfield  {journal} {\bibinfo
   {journal} {Phys. Rev. Lett.}\ }\textbf {\bibinfo {volume} {72}},\ \bibinfo
  {pages} {1790} (\bibinfo {year} {1994})},\ \Eprint
  {http://arxiv.org/abs/hep-ph/9306299} {arXiv:hep-ph/9306299} \BibitemShut
  {NoStop}%
\bibitem [{\citenamefont {Detmold}\ and\ \citenamefont
  {Lin}(2006)}]{Detmold:2005gg}%
  \BibitemOpen
  \bibfield  {author} {\bibinfo {author} {\bibfnamefont {W.}~\bibnamefont
  {Detmold}}\ and\ \bibinfo {author} {\bibfnamefont {C.~J.~D.}\ \bibnamefont
  {Lin}},\ }\href {\doibase 10.1103/PhysRevD.73.014501} {\bibfield  {journal}
  {\bibinfo  {journal} {Phys. Rev. D}\ }\textbf {\bibinfo {volume} {73}},\
  \bibinfo {pages} {014501} (\bibinfo {year} {2006})},\ \Eprint
  {http://arxiv.org/abs/hep-lat/0507007} {arXiv:hep-lat/0507007} \BibitemShut
  {NoStop}%
\bibitem [{\citenamefont {Braun}\ and\ \citenamefont
  {M\"uller}(2008)}]{Braun:2007wv}%
  \BibitemOpen
  \bibfield  {author} {\bibinfo {author} {\bibfnamefont {V.}~\bibnamefont
  {Braun}}\ and\ \bibinfo {author} {\bibfnamefont {D.}~\bibnamefont
  {M\"uller}},\ }\href {\doibase 10.1140/epjc/s10052-008-0608-4} {\bibfield
  {journal} {\bibinfo  {journal} {Eur. Phys. J. C}\ }\textbf {\bibinfo {volume}
  {55}},\ \bibinfo {pages} {349} (\bibinfo {year} {2008})},\ \Eprint
  {http://arxiv.org/abs/0709.1348} {arXiv:0709.1348 [hep-ph]} \BibitemShut
  {NoStop}%
\bibitem [{\citenamefont {Ji}(2013)}]{Ji:2013dva}%
  \BibitemOpen
  \bibfield  {author} {\bibinfo {author} {\bibfnamefont {X.}~\bibnamefont
  {Ji}},\ }\href {\doibase 10.1103/PhysRevLett.110.262002} {\bibfield
  {journal} {\bibinfo  {journal} {Phys. Rev. Lett.}\ }\textbf {\bibinfo
  {volume} {110}},\ \bibinfo {pages} {262002} (\bibinfo {year} {2013})},\
  \Eprint {http://arxiv.org/abs/1305.1539} {arXiv:1305.1539 [hep-ph]}
  \BibitemShut {NoStop}%
\bibitem [{\citenamefont {Ji}(2014)}]{Ji:2014gla}%
  \BibitemOpen
  \bibfield  {author} {\bibinfo {author} {\bibfnamefont {X.}~\bibnamefont
  {Ji}},\ }\href {\doibase 10.1007/s11433-014-5492-3} {\bibfield  {journal}
  {\bibinfo  {journal} {Sci. China Phys. Mech. Astron.}\ }\textbf {\bibinfo
  {volume} {57}},\ \bibinfo {pages} {1407} (\bibinfo {year} {2014})},\ \Eprint
  {http://arxiv.org/abs/1404.6680} {arXiv:1404.6680 [hep-ph]} \BibitemShut
  {NoStop}%
\bibitem [{\citenamefont {Chambers}\ \emph {et~al.}(2017)\citenamefont
  {Chambers}, \citenamefont {Horsley}, \citenamefont {Nakamura}, \citenamefont
  {Perlt}, \citenamefont {Rakow}, \citenamefont {Schierholz}, \citenamefont
  {Schiller}, \citenamefont {Somfleth}, \citenamefont {Young},\ and\
  \citenamefont {Zanotti}}]{Chambers:2017dov}%
  \BibitemOpen
  \bibfield  {author} {\bibinfo {author} {\bibfnamefont {A.~J.}\ \bibnamefont
  {Chambers}}, \bibinfo {author} {\bibfnamefont {R.}~\bibnamefont {Horsley}},
  \bibinfo {author} {\bibfnamefont {Y.}~\bibnamefont {Nakamura}}, \bibinfo
  {author} {\bibfnamefont {H.}~\bibnamefont {Perlt}}, \bibinfo {author}
  {\bibfnamefont {P.~E.~L.}\ \bibnamefont {Rakow}}, \bibinfo {author}
  {\bibfnamefont {G.}~\bibnamefont {Schierholz}}, \bibinfo {author}
  {\bibfnamefont {A.}~\bibnamefont {Schiller}}, \bibinfo {author}
  {\bibfnamefont {K.}~\bibnamefont {Somfleth}}, \bibinfo {author}
  {\bibfnamefont {R.~D.}\ \bibnamefont {Young}}, \ and\ \bibinfo {author}
  {\bibfnamefont {J.~M.}\ \bibnamefont {Zanotti}},\ }\href {\doibase
  10.1103/PhysRevLett.118.242001} {\bibfield  {journal} {\bibinfo  {journal}
  {Phys. Rev. Lett.}\ }\textbf {\bibinfo {volume} {118}},\ \bibinfo {pages}
  {242001} (\bibinfo {year} {2017})},\ \Eprint
  {http://arxiv.org/abs/1703.01153} {arXiv:1703.01153 [hep-lat]} \BibitemShut
  {NoStop}%
\bibitem [{\citenamefont {Radyushkin}(2017)}]{Radyushkin:2017cyf}%
  \BibitemOpen
  \bibfield  {author} {\bibinfo {author} {\bibfnamefont {A.~V.}\ \bibnamefont
  {Radyushkin}},\ }\href {\doibase 10.1103/PhysRevD.96.034025} {\bibfield
  {journal} {\bibinfo  {journal} {Phys. Rev. D}\ }\textbf {\bibinfo {volume}
  {96}},\ \bibinfo {pages} {034025} (\bibinfo {year} {2017})},\ \Eprint
  {http://arxiv.org/abs/1705.01488} {arXiv:1705.01488 [hep-ph]} \BibitemShut
  {NoStop}%
\bibitem [{\citenamefont {Ma}\ and\ \citenamefont
  {Qiu}(2018{\natexlab{a}})}]{Ma:2014jla}%
  \BibitemOpen
  \bibfield  {author} {\bibinfo {author} {\bibfnamefont {Y.-Q.}\ \bibnamefont
  {Ma}}\ and\ \bibinfo {author} {\bibfnamefont {J.-W.}\ \bibnamefont {Qiu}},\
  }\href {\doibase 10.1103/PhysRevD.98.074021} {\bibfield  {journal} {\bibinfo
  {journal} {Phys. Rev. D}\ }\textbf {\bibinfo {volume} {98}},\ \bibinfo
  {pages} {074021} (\bibinfo {year} {2018}{\natexlab{a}})},\ \Eprint
  {http://arxiv.org/abs/1404.6860} {arXiv:1404.6860 [hep-ph]} \BibitemShut
  {NoStop}%
\bibitem [{\citenamefont {Ma}\ and\ \citenamefont
  {Qiu}(2018{\natexlab{b}})}]{Ma:2017pxb}%
  \BibitemOpen
  \bibfield  {author} {\bibinfo {author} {\bibfnamefont {Y.-Q.}\ \bibnamefont
  {Ma}}\ and\ \bibinfo {author} {\bibfnamefont {J.-W.}\ \bibnamefont {Qiu}},\
  }\href {\doibase 10.1103/PhysRevLett.120.022003} {\bibfield  {journal}
  {\bibinfo  {journal} {Phys. Rev. Lett.}\ }\textbf {\bibinfo {volume} {120}},\
  \bibinfo {pages} {022003} (\bibinfo {year} {2018}{\natexlab{b}})},\ \Eprint
  {http://arxiv.org/abs/1709.03018} {arXiv:1709.03018 [hep-ph]} \BibitemShut
  {NoStop}%
\bibitem [{\citenamefont {Cichy}\ and\ \citenamefont
  {Constantinou}(2019)}]{Cichy:2018mum}%
  \BibitemOpen
  \bibfield  {author} {\bibinfo {author} {\bibfnamefont {K.}~\bibnamefont
  {Cichy}}\ and\ \bibinfo {author} {\bibfnamefont {M.}~\bibnamefont
  {Constantinou}},\ }\href {\doibase 10.1155/2019/3036904} {\bibfield
  {journal} {\bibinfo  {journal} {Adv. High Energy Phys.}\ }\textbf {\bibinfo
  {volume} {2019}},\ \bibinfo {pages} {3036904} (\bibinfo {year} {2019})},\
  \Eprint {http://arxiv.org/abs/1811.07248} {arXiv:1811.07248 [hep-lat]}
  \BibitemShut {NoStop}%
\bibitem [{\citenamefont {Constantinou}\ \emph {et~al.}(2021)\citenamefont
  {Constantinou} \emph {et~al.}}]{Constantinou:2020hdm}%
  \BibitemOpen
  \bibfield  {author} {\bibinfo {author} {\bibfnamefont {M.}~\bibnamefont
  {Constantinou}} \emph {et~al.},\ }\href {\doibase 10.1016/j.ppnp.2021.103908}
  {\bibfield  {journal} {\bibinfo  {journal} {Prog. Part. Nucl. Phys.}\
  }\textbf {\bibinfo {volume} {121}},\ \bibinfo {pages} {103908} (\bibinfo
  {year} {2021})},\ \Eprint {http://arxiv.org/abs/2006.08636} {arXiv:2006.08636
  [hep-ph]} \BibitemShut {NoStop}%
\bibitem [{\citenamefont {Ji}\ \emph {et~al.}(2021{\natexlab{b}})\citenamefont
  {Ji}, \citenamefont {Liu}, \citenamefont {Liu}, \citenamefont {Zhang},\ and\
  \citenamefont {Zhao}}]{Ji:2020ect}%
  \BibitemOpen
  \bibfield  {author} {\bibinfo {author} {\bibfnamefont {X.}~\bibnamefont
  {Ji}}, \bibinfo {author} {\bibfnamefont {Y.-S.}\ \bibnamefont {Liu}},
  \bibinfo {author} {\bibfnamefont {Y.}~\bibnamefont {Liu}}, \bibinfo {author}
  {\bibfnamefont {J.-H.}\ \bibnamefont {Zhang}}, \ and\ \bibinfo {author}
  {\bibfnamefont {Y.}~\bibnamefont {Zhao}},\ }\href {\doibase
  10.1103/RevModPhys.93.035005} {\bibfield  {journal} {\bibinfo  {journal}
  {Rev. Mod. Phys.}\ }\textbf {\bibinfo {volume} {93}},\ \bibinfo {pages}
  {035005} (\bibinfo {year} {2021}{\natexlab{b}})},\ \Eprint
  {http://arxiv.org/abs/2004.03543} {arXiv:2004.03543 [hep-ph]} \BibitemShut
  {NoStop}%
\bibitem [{\citenamefont {Constantinou}\ \emph {et~al.}(2022)\citenamefont
  {Constantinou} \emph {et~al.}}]{Constantinou:2022yye}%
  \BibitemOpen
  \bibfield  {author} {\bibinfo {author} {\bibfnamefont {M.}~\bibnamefont
  {Constantinou}} \emph {et~al.},\ }\href@noop {} {\  (\bibinfo {year}
  {2022})},\ \Eprint {http://arxiv.org/abs/2202.07193} {arXiv:2202.07193
  [hep-lat]} \BibitemShut {NoStop}%
\bibitem [{\citenamefont {Zhang}\ \emph {et~al.}(2019)\citenamefont {Zhang},
  \citenamefont {Ji}, \citenamefont {Sch\"afer}, \citenamefont {Wang},\ and\
  \citenamefont {Zhao}}]{Zhang:2018diq}%
  \BibitemOpen
  \bibfield  {author} {\bibinfo {author} {\bibfnamefont {J.-H.}\ \bibnamefont
  {Zhang}}, \bibinfo {author} {\bibfnamefont {X.}~\bibnamefont {Ji}}, \bibinfo
  {author} {\bibfnamefont {A.}~\bibnamefont {Sch\"afer}}, \bibinfo {author}
  {\bibfnamefont {W.}~\bibnamefont {Wang}}, \ and\ \bibinfo {author}
  {\bibfnamefont {S.}~\bibnamefont {Zhao}},\ }\href {\doibase
  10.1103/PhysRevLett.122.142001} {\bibfield  {journal} {\bibinfo  {journal}
  {Phys. Rev. Lett.}\ }\textbf {\bibinfo {volume} {122}},\ \bibinfo {pages}
  {142001} (\bibinfo {year} {2019})},\ \Eprint
  {http://arxiv.org/abs/1808.10824} {arXiv:1808.10824 [hep-ph]} \BibitemShut
  {NoStop}%
\bibitem [{\citenamefont {Li}\ \emph {et~al.}(2019)\citenamefont {Li},
  \citenamefont {Ma},\ and\ \citenamefont {Qiu}}]{Li:2018tpe}%
  \BibitemOpen
  \bibfield  {author} {\bibinfo {author} {\bibfnamefont {Z.-Y.}\ \bibnamefont
  {Li}}, \bibinfo {author} {\bibfnamefont {Y.-Q.}\ \bibnamefont {Ma}}, \ and\
  \bibinfo {author} {\bibfnamefont {J.-W.}\ \bibnamefont {Qiu}},\ }\href
  {\doibase 10.1103/PhysRevLett.122.062002} {\bibfield  {journal} {\bibinfo
  {journal} {Phys. Rev. Lett.}\ }\textbf {\bibinfo {volume} {122}},\ \bibinfo
  {pages} {062002} (\bibinfo {year} {2019})},\ \Eprint
  {http://arxiv.org/abs/1809.01836} {arXiv:1809.01836 [hep-ph]} \BibitemShut
  {NoStop}%
\bibitem [{\citenamefont {Izubuchi}\ \emph {et~al.}(2018)\citenamefont
  {Izubuchi}, \citenamefont {Ji}, \citenamefont {Jin}, \citenamefont
  {Stewart},\ and\ \citenamefont {Zhao}}]{Izubuchi:2018srq}%
  \BibitemOpen
  \bibfield  {author} {\bibinfo {author} {\bibfnamefont {T.}~\bibnamefont
  {Izubuchi}}, \bibinfo {author} {\bibfnamefont {X.}~\bibnamefont {Ji}},
  \bibinfo {author} {\bibfnamefont {L.}~\bibnamefont {Jin}}, \bibinfo {author}
  {\bibfnamefont {I.~W.}\ \bibnamefont {Stewart}}, \ and\ \bibinfo {author}
  {\bibfnamefont {Y.}~\bibnamefont {Zhao}},\ }\href {\doibase
  10.1103/PhysRevD.98.056004} {\bibfield  {journal} {\bibinfo  {journal} {Phys.
  Rev. D}\ }\textbf {\bibinfo {volume} {98}},\ \bibinfo {pages} {056004}
  (\bibinfo {year} {2018})},\ \Eprint {http://arxiv.org/abs/1801.03917}
  {arXiv:1801.03917 [hep-ph]} \BibitemShut {NoStop}%
\bibitem [{\citenamefont {Ji}\ \emph {et~al.}(2018)\citenamefont {Ji},
  \citenamefont {Zhang},\ and\ \citenamefont {Zhao}}]{Ji:2017oey}%
  \BibitemOpen
  \bibfield  {author} {\bibinfo {author} {\bibfnamefont {X.}~\bibnamefont
  {Ji}}, \bibinfo {author} {\bibfnamefont {J.-H.}\ \bibnamefont {Zhang}}, \
  and\ \bibinfo {author} {\bibfnamefont {Y.}~\bibnamefont {Zhao}},\ }\href
  {\doibase 10.1103/PhysRevLett.120.112001} {\bibfield  {journal} {\bibinfo
  {journal} {Phys. Rev. Lett.}\ }\textbf {\bibinfo {volume} {120}},\ \bibinfo
  {pages} {112001} (\bibinfo {year} {2018})},\ \Eprint
  {http://arxiv.org/abs/1706.08962} {arXiv:1706.08962 [hep-ph]} \BibitemShut
  {NoStop}%
\bibitem [{\citenamefont {Green}\ \emph {et~al.}(2018)\citenamefont {Green},
  \citenamefont {Jansen},\ and\ \citenamefont {Steffens}}]{Green:2017xeu}%
  \BibitemOpen
  \bibfield  {author} {\bibinfo {author} {\bibfnamefont {J.}~\bibnamefont
  {Green}}, \bibinfo {author} {\bibfnamefont {K.}~\bibnamefont {Jansen}}, \
  and\ \bibinfo {author} {\bibfnamefont {F.}~\bibnamefont {Steffens}},\ }\href
  {\doibase 10.1103/PhysRevLett.121.022004} {\bibfield  {journal} {\bibinfo
  {journal} {Phys. Rev. Lett.}\ }\textbf {\bibinfo {volume} {121}},\ \bibinfo
  {pages} {022004} (\bibinfo {year} {2018})},\ \Eprint
  {http://arxiv.org/abs/1707.07152} {arXiv:1707.07152 [hep-lat]} \BibitemShut
  {NoStop}%
\bibitem [{\citenamefont {Gribov}\ \emph {et~al.}(1965)\citenamefont {Gribov},
  \citenamefont {Ioffe},\ and\ \citenamefont {Pomeranchuk}}]{Gribov:1965hf}%
  \BibitemOpen
  \bibfield  {author} {\bibinfo {author} {\bibfnamefont {V.~N.}\ \bibnamefont
  {Gribov}}, \bibinfo {author} {\bibfnamefont {B.~L.}\ \bibnamefont {Ioffe}}, \
  and\ \bibinfo {author} {\bibfnamefont {I.~Y.}\ \bibnamefont {Pomeranchuk}},\
  }\href@noop {} {\bibfield  {journal} {\bibinfo  {journal} {Yad. Fiz.}\
  }\textbf {\bibinfo {volume} {2}},\ \bibinfo {pages} {768} (\bibinfo {year}
  {1965})}\BibitemShut {NoStop}%
\bibitem [{\citenamefont {Ioffe}(1969)}]{Ioffe:1969kf}%
  \BibitemOpen
  \bibfield  {author} {\bibinfo {author} {\bibfnamefont {B.~L.}\ \bibnamefont
  {Ioffe}},\ }\href {\doibase 10.1016/0370-2693(69)90415-8} {\bibfield
  {journal} {\bibinfo  {journal} {Phys. Lett. B}\ }\textbf {\bibinfo {volume}
  {30}},\ \bibinfo {pages} {123} (\bibinfo {year} {1969})}\BibitemShut
  {NoStop}%
\bibitem [{\citenamefont {Braun}\ \emph {et~al.}(1995)\citenamefont {Braun},
  \citenamefont {Gornicki},\ and\ \citenamefont {Mankiewicz}}]{Braun:1994jq}%
  \BibitemOpen
  \bibfield  {author} {\bibinfo {author} {\bibfnamefont {V.}~\bibnamefont
  {Braun}}, \bibinfo {author} {\bibfnamefont {P.}~\bibnamefont {Gornicki}}, \
  and\ \bibinfo {author} {\bibfnamefont {L.}~\bibnamefont {Mankiewicz}},\
  }\href {\doibase 10.1103/PhysRevD.51.6036} {\bibfield  {journal} {\bibinfo
  {journal} {Phys. Rev. D}\ }\textbf {\bibinfo {volume} {51}},\ \bibinfo
  {pages} {6036} (\bibinfo {year} {1995})},\ \Eprint
  {http://arxiv.org/abs/hep-ph/9410318} {arXiv:hep-ph/9410318} \BibitemShut
  {NoStop}%
\bibitem [{\citenamefont {Orginos}\ \emph {et~al.}(2017)\citenamefont
  {Orginos}, \citenamefont {Radyushkin}, \citenamefont {Karpie},\ and\
  \citenamefont {Zafeiropoulos}}]{Orginos:2017kos}%
  \BibitemOpen
  \bibfield  {author} {\bibinfo {author} {\bibfnamefont {K.}~\bibnamefont
  {Orginos}}, \bibinfo {author} {\bibfnamefont {A.}~\bibnamefont {Radyushkin}},
  \bibinfo {author} {\bibfnamefont {J.}~\bibnamefont {Karpie}}, \ and\ \bibinfo
  {author} {\bibfnamefont {S.}~\bibnamefont {Zafeiropoulos}},\ }\href {\doibase
  10.1103/PhysRevD.96.094503} {\bibfield  {journal} {\bibinfo  {journal} {Phys.
  Rev. D}\ }\textbf {\bibinfo {volume} {96}},\ \bibinfo {pages} {094503}
  (\bibinfo {year} {2017})},\ \Eprint {http://arxiv.org/abs/1706.05373}
  {arXiv:1706.05373 [hep-ph]} \BibitemShut {NoStop}%
\bibitem [{\citenamefont {Fan}\ \emph {et~al.}(2018)\citenamefont {Fan},
  \citenamefont {Yang}, \citenamefont {Anthony}, \citenamefont {Lin},\ and\
  \citenamefont {Liu}}]{Fan:2018dxu}%
  \BibitemOpen
  \bibfield  {author} {\bibinfo {author} {\bibfnamefont {Z.-Y.}\ \bibnamefont
  {Fan}}, \bibinfo {author} {\bibfnamefont {Y.-B.}\ \bibnamefont {Yang}},
  \bibinfo {author} {\bibfnamefont {A.}~\bibnamefont {Anthony}}, \bibinfo
  {author} {\bibfnamefont {H.-W.}\ \bibnamefont {Lin}}, \ and\ \bibinfo
  {author} {\bibfnamefont {K.-F.}\ \bibnamefont {Liu}},\ }\href {\doibase
  10.1103/PhysRevLett.121.242001} {\bibfield  {journal} {\bibinfo  {journal}
  {Phys. Rev. Lett.}\ }\textbf {\bibinfo {volume} {121}},\ \bibinfo {pages}
  {242001} (\bibinfo {year} {2018})},\ \Eprint
  {http://arxiv.org/abs/1808.02077} {arXiv:1808.02077 [hep-lat]} \BibitemShut
  {NoStop}%
\bibitem [{\citenamefont {Fan}\ \emph {et~al.}(2021)\citenamefont {Fan},
  \citenamefont {Zhang},\ and\ \citenamefont {Lin}}]{Fan:2020cpa}%
  \BibitemOpen
  \bibfield  {author} {\bibinfo {author} {\bibfnamefont {Z.}~\bibnamefont
  {Fan}}, \bibinfo {author} {\bibfnamefont {R.}~\bibnamefont {Zhang}}, \ and\
  \bibinfo {author} {\bibfnamefont {H.-W.}\ \bibnamefont {Lin}},\ }\href
  {\doibase 10.1142/S0217751X21500809} {\bibfield  {journal} {\bibinfo
  {journal} {Int. J. Mod. Phys. A}\ }\textbf {\bibinfo {volume} {36}},\
  \bibinfo {pages} {2150080} (\bibinfo {year} {2021})},\ \Eprint
  {http://arxiv.org/abs/2007.16113} {arXiv:2007.16113 [hep-lat]} \BibitemShut
  {NoStop}%
\bibitem [{\citenamefont {Khan}\ \emph {et~al.}(2021)\citenamefont {Khan} \emph
  {et~al.}}]{HadStruc:2021wmh}%
  \BibitemOpen
  \bibfield  {author} {\bibinfo {author} {\bibfnamefont {T.}~\bibnamefont
  {Khan}} \emph {et~al.} (\bibinfo {collaboration} {HadStruc Collaboration}),\
  }\href {\doibase 10.1103/PhysRevD.104.094516} {\bibfield  {journal} {\bibinfo
   {journal} {Phys. Rev. D}\ }\textbf {\bibinfo {volume} {104}},\ \bibinfo
  {pages} {094516} (\bibinfo {year} {2021})},\ \Eprint
  {http://arxiv.org/abs/2107.08960} {arXiv:2107.08960 [hep-lat]} \BibitemShut
  {NoStop}%
\bibitem [{\citenamefont {Fan}\ and\ \citenamefont {Lin}(2021)}]{Fan:2021bcr}%
  \BibitemOpen
  \bibfield  {author} {\bibinfo {author} {\bibfnamefont {Z.}~\bibnamefont
  {Fan}}\ and\ \bibinfo {author} {\bibfnamefont {H.-W.}\ \bibnamefont {Lin}},\
  }\href {\doibase 10.1016/j.physletb.2021.136778} {\bibfield  {journal}
  {\bibinfo  {journal} {Phys. Lett. B}\ }\textbf {\bibinfo {volume} {823}},\
  \bibinfo {pages} {136778} (\bibinfo {year} {2021})},\ \Eprint
  {http://arxiv.org/abs/2104.06372} {arXiv:2104.06372 [hep-lat]} \BibitemShut
  {NoStop}%
\bibitem [{\citenamefont {Salas-Chavira}\ \emph {et~al.}(2021)\citenamefont
  {Salas-Chavira}, \citenamefont {Fan},\ and\ \citenamefont
  {Lin}}]{Salas-Chavira:2021wui}%
  \BibitemOpen
  \bibfield  {author} {\bibinfo {author} {\bibfnamefont {A.}~\bibnamefont
  {Salas-Chavira}}, \bibinfo {author} {\bibfnamefont {Z.}~\bibnamefont {Fan}},
  \ and\ \bibinfo {author} {\bibfnamefont {H.-W.}\ \bibnamefont {Lin}},\
  }\href@noop {} {\  (\bibinfo {year} {2021})},\ \Eprint
  {http://arxiv.org/abs/2112.03124} {arXiv:2112.03124 [hep-lat]} \BibitemShut
  {NoStop}%
\bibitem [{\citenamefont {Fan}\ \emph {et~al.}(2022)\citenamefont {Fan},
  \citenamefont {Good},\ and\ \citenamefont {Lin}}]{Fan:2022kcb}%
  \BibitemOpen
  \bibfield  {author} {\bibinfo {author} {\bibfnamefont {Z.}~\bibnamefont
  {Fan}}, \bibinfo {author} {\bibfnamefont {W.}~\bibnamefont {Good}}, \ and\
  \bibinfo {author} {\bibfnamefont {H.-W.}\ \bibnamefont {Lin}},\ }\href@noop
  {} {\  (\bibinfo {year} {2022})},\ \Eprint {http://arxiv.org/abs/2210.09985}
  {arXiv:2210.09985 [hep-lat]} \BibitemShut {NoStop}%
\bibitem [{\citenamefont {Egerer}\ \emph {et~al.}(2022)\citenamefont {Egerer}
  \emph {et~al.}}]{HadStruc:2022yaw}%
  \BibitemOpen
  \bibfield  {author} {\bibinfo {author} {\bibfnamefont {C.}~\bibnamefont
  {Egerer}} \emph {et~al.} (\bibinfo {collaboration} {HadStruc}),\ }\href
  {\doibase 10.1103/PhysRevD.106.094511} {\bibfield  {journal} {\bibinfo
  {journal} {Phys. Rev. D}\ }\textbf {\bibinfo {volume} {106}},\ \bibinfo
  {pages} {094511} (\bibinfo {year} {2022})},\ \Eprint
  {http://arxiv.org/abs/2207.08733} {arXiv:2207.08733 [hep-lat]} \BibitemShut
  {NoStop}%
\bibitem [{\citenamefont {Balitsky}\ \emph {et~al.}(2022)\citenamefont
  {Balitsky}, \citenamefont {Morris},\ and\ \citenamefont
  {Radyushkin}}]{Balitsky:2021cwr}%
  \BibitemOpen
  \bibfield  {author} {\bibinfo {author} {\bibfnamefont {I.}~\bibnamefont
  {Balitsky}}, \bibinfo {author} {\bibfnamefont {W.}~\bibnamefont {Morris}}, \
  and\ \bibinfo {author} {\bibfnamefont {A.}~\bibnamefont {Radyushkin}},\
  }\href {\doibase 10.1007/JHEP02(2022)193} {\bibfield  {journal} {\bibinfo
  {journal} {JHEP}\ }\textbf {\bibinfo {volume} {02}},\ \bibinfo {pages} {193}
  (\bibinfo {year} {2022})},\ \Eprint {http://arxiv.org/abs/2112.02011}
  {arXiv:2112.02011 [hep-ph]} \BibitemShut {NoStop}%
\bibitem [{\citenamefont {Balitsky}\ \emph {et~al.}(2020)\citenamefont
  {Balitsky}, \citenamefont {Morris},\ and\ \citenamefont
  {Radyushkin}}]{Balitsky:2019krf}%
  \BibitemOpen
  \bibfield  {author} {\bibinfo {author} {\bibfnamefont {I.}~\bibnamefont
  {Balitsky}}, \bibinfo {author} {\bibfnamefont {W.}~\bibnamefont {Morris}}, \
  and\ \bibinfo {author} {\bibfnamefont {A.}~\bibnamefont {Radyushkin}},\
  }\href {\doibase 10.1016/j.physletb.2020.135621} {\bibfield  {journal}
  {\bibinfo  {journal} {Phys. Lett. B}\ }\textbf {\bibinfo {volume} {808}},\
  \bibinfo {pages} {135621} (\bibinfo {year} {2020})},\ \Eprint
  {http://arxiv.org/abs/1910.13963} {arXiv:1910.13963 [hep-ph]} \BibitemShut
  {NoStop}%
\bibitem [{\citenamefont {Ji}\ \emph {et~al.}(2021{\natexlab{c}})\citenamefont
  {Ji}, \citenamefont {Liu}, \citenamefont {Sch\"afer}, \citenamefont {Wang},
  \citenamefont {Yang}, \citenamefont {Zhang},\ and\ \citenamefont
  {Zhao}}]{Ji:2020brr}%
  \BibitemOpen
  \bibfield  {author} {\bibinfo {author} {\bibfnamefont {X.}~\bibnamefont
  {Ji}}, \bibinfo {author} {\bibfnamefont {Y.}~\bibnamefont {Liu}}, \bibinfo
  {author} {\bibfnamefont {A.}~\bibnamefont {Sch\"afer}}, \bibinfo {author}
  {\bibfnamefont {W.}~\bibnamefont {Wang}}, \bibinfo {author} {\bibfnamefont
  {Y.-B.}\ \bibnamefont {Yang}}, \bibinfo {author} {\bibfnamefont {J.-H.}\
  \bibnamefont {Zhang}}, \ and\ \bibinfo {author} {\bibfnamefont
  {Y.}~\bibnamefont {Zhao}},\ }\href {\doibase 10.1016/j.nuclphysb.2021.115311}
  {\bibfield  {journal} {\bibinfo  {journal} {Nucl. Phys. B}\ }\textbf
  {\bibinfo {volume} {964}},\ \bibinfo {pages} {115311} (\bibinfo {year}
  {2021}{\natexlab{c}})},\ \Eprint {http://arxiv.org/abs/2008.03886}
  {arXiv:2008.03886 [hep-ph]} \BibitemShut {NoStop}%
\bibitem [{\citenamefont {Sufian}\ \emph {et~al.}(2021)\citenamefont {Sufian},
  \citenamefont {Liu},\ and\ \citenamefont {Paul}}]{Sufian:2020wcv}%
  \BibitemOpen
  \bibfield  {author} {\bibinfo {author} {\bibfnamefont {R.~S.}\ \bibnamefont
  {Sufian}}, \bibinfo {author} {\bibfnamefont {T.}~\bibnamefont {Liu}}, \ and\
  \bibinfo {author} {\bibfnamefont {A.}~\bibnamefont {Paul}},\ }\href {\doibase
  10.1103/PhysRevD.103.036007} {\bibfield  {journal} {\bibinfo  {journal}
  {Phys. Rev. D}\ }\textbf {\bibinfo {volume} {103}},\ \bibinfo {pages}
  {036007} (\bibinfo {year} {2021})},\ \Eprint
  {http://arxiv.org/abs/2012.01532} {arXiv:2012.01532 [hep-ph]} \BibitemShut
  {NoStop}%
\bibitem [{\citenamefont {Saalfeld}\ \emph {et~al.}(1998)\citenamefont
  {Saalfeld}, \citenamefont {Piller},\ and\ \citenamefont
  {Mankiewicz}}]{Saalfeld:1997uv}%
  \BibitemOpen
  \bibfield  {author} {\bibinfo {author} {\bibfnamefont {A.}~\bibnamefont
  {Saalfeld}}, \bibinfo {author} {\bibfnamefont {G.}~\bibnamefont {Piller}}, \
  and\ \bibinfo {author} {\bibfnamefont {L.}~\bibnamefont {Mankiewicz}},\
  }\href {\doibase 10.1007/s100520050209} {\bibfield  {journal} {\bibinfo
  {journal} {Eur. Phys. J. C}\ }\textbf {\bibinfo {volume} {4}},\ \bibinfo
  {pages} {307} (\bibinfo {year} {1998})},\ \Eprint
  {http://arxiv.org/abs/hep-ph/9708378} {arXiv:hep-ph/9708378} \BibitemShut
  {NoStop}%
\bibitem [{\citenamefont {Egerer}\ \emph {et~al.}(2021)\citenamefont {Egerer},
  \citenamefont {Edwards}, \citenamefont {Kallidonis}, \citenamefont {Orginos},
  \citenamefont {Radyushkin}, \citenamefont {Richards}, \citenamefont
  {Romero},\ and\ \citenamefont {Zafeiropoulos}}]{Egerer:2021ymv}%
  \BibitemOpen
  \bibfield  {author} {\bibinfo {author} {\bibfnamefont {C.}~\bibnamefont
  {Egerer}}, \bibinfo {author} {\bibfnamefont {R.~G.}\ \bibnamefont {Edwards}},
  \bibinfo {author} {\bibfnamefont {C.}~\bibnamefont {Kallidonis}}, \bibinfo
  {author} {\bibfnamefont {K.}~\bibnamefont {Orginos}}, \bibinfo {author}
  {\bibfnamefont {A.~V.}\ \bibnamefont {Radyushkin}}, \bibinfo {author}
  {\bibfnamefont {D.~G.}\ \bibnamefont {Richards}}, \bibinfo {author}
  {\bibfnamefont {E.}~\bibnamefont {Romero}}, \ and\ \bibinfo {author}
  {\bibfnamefont {S.}~\bibnamefont {Zafeiropoulos}} (\bibinfo {collaboration}
  {HadStruc Collaboration}),\ }\href {\doibase 10.1007/JHEP11(2021)148}
  {\bibfield  {journal} {\bibinfo  {journal} {JHEP}\ }\textbf {\bibinfo
  {volume} {11}},\ \bibinfo {pages} {148} (\bibinfo {year} {2021})},\ \Eprint
  {http://arxiv.org/abs/2107.05199} {arXiv:2107.05199 [hep-lat]} \BibitemShut
  {NoStop}%
\bibitem [{\citenamefont {Bhat}\ \emph {et~al.}(2022)\citenamefont {Bhat},
  \citenamefont {Chomicki}, \citenamefont {Cichy}, \citenamefont
  {Constantinou}, \citenamefont {Green},\ and\ \citenamefont
  {Scapellato}}]{Bhat:2022zrw}%
  \BibitemOpen
  \bibfield  {author} {\bibinfo {author} {\bibfnamefont {M.}~\bibnamefont
  {Bhat}}, \bibinfo {author} {\bibfnamefont {W.}~\bibnamefont {Chomicki}},
  \bibinfo {author} {\bibfnamefont {K.}~\bibnamefont {Cichy}}, \bibinfo
  {author} {\bibfnamefont {M.}~\bibnamefont {Constantinou}}, \bibinfo {author}
  {\bibfnamefont {J.~R.}\ \bibnamefont {Green}}, \ and\ \bibinfo {author}
  {\bibfnamefont {A.}~\bibnamefont {Scapellato}},\ }\href {\doibase
  10.1103/PhysRevD.106.054504} {\bibfield  {journal} {\bibinfo  {journal}
  {Phys. Rev. D}\ }\textbf {\bibinfo {volume} {106}},\ \bibinfo {pages}
  {054504} (\bibinfo {year} {2022})},\ \Eprint
  {http://arxiv.org/abs/2205.07585} {arXiv:2205.07585 [hep-lat]} \BibitemShut
  {NoStop}%
\bibitem [{\citenamefont {Li}\ \emph {et~al.}(2021)\citenamefont {Li},
  \citenamefont {Ma},\ and\ \citenamefont {Qiu}}]{Li:2020xml}%
  \BibitemOpen
  \bibfield  {author} {\bibinfo {author} {\bibfnamefont {Z.-Y.}\ \bibnamefont
  {Li}}, \bibinfo {author} {\bibfnamefont {Y.-Q.}\ \bibnamefont {Ma}}, \ and\
  \bibinfo {author} {\bibfnamefont {J.-W.}\ \bibnamefont {Qiu}},\ }\href
  {\doibase 10.1103/PhysRevLett.126.072001} {\bibfield  {journal} {\bibinfo
  {journal} {Phys. Rev. Lett.}\ }\textbf {\bibinfo {volume} {126}},\ \bibinfo
  {pages} {072001} (\bibinfo {year} {2021})},\ \Eprint
  {http://arxiv.org/abs/2006.12370} {arXiv:2006.12370 [hep-ph]} \BibitemShut
  {NoStop}%
\bibitem [{\citenamefont {Karthik}\ and\ \citenamefont
  {Sufian}(2021)}]{Karthik:2021sbj}%
  \BibitemOpen
  \bibfield  {author} {\bibinfo {author} {\bibfnamefont {N.}~\bibnamefont
  {Karthik}}\ and\ \bibinfo {author} {\bibfnamefont {R.~S.}\ \bibnamefont
  {Sufian}},\ }\href {\doibase 10.1103/PhysRevD.104.074506} {\bibfield
  {journal} {\bibinfo  {journal} {Phys. Rev. D}\ }\textbf {\bibinfo {volume}
  {104}},\ \bibinfo {pages} {074506} (\bibinfo {year} {2021})},\ \Eprint
  {http://arxiv.org/abs/2106.03875} {arXiv:2106.03875 [hep-lat]} \BibitemShut
  {NoStop}%
\bibitem [{\citenamefont {Ji}(2022)}]{Ji:2022ezo}%
  \BibitemOpen
  \bibfield  {author} {\bibinfo {author} {\bibfnamefont {X.}~\bibnamefont
  {Ji}},\ }\href@noop {} {\  (\bibinfo {year} {2022})},\ \Eprint
  {http://arxiv.org/abs/2209.09332} {arXiv:2209.09332 [hep-lat]} \BibitemShut
  {NoStop}%
\bibitem [{\citenamefont {Su}\ \emph {et~al.}(2022)\citenamefont {Su},
  \citenamefont {Holligan}, \citenamefont {Ji}, \citenamefont {Yao},
  \citenamefont {Zhang},\ and\ \citenamefont {Zhang}}]{Su:2022fiu}%
  \BibitemOpen
  \bibfield  {author} {\bibinfo {author} {\bibfnamefont {Y.}~\bibnamefont
  {Su}}, \bibinfo {author} {\bibfnamefont {J.}~\bibnamefont {Holligan}},
  \bibinfo {author} {\bibfnamefont {X.}~\bibnamefont {Ji}}, \bibinfo {author}
  {\bibfnamefont {F.}~\bibnamefont {Yao}}, \bibinfo {author} {\bibfnamefont
  {J.-H.}\ \bibnamefont {Zhang}}, \ and\ \bibinfo {author} {\bibfnamefont
  {R.}~\bibnamefont {Zhang}},\ }\href@noop {} {\  (\bibinfo {year} {2022})},\
  \Eprint {http://arxiv.org/abs/2209.01236} {arXiv:2209.01236 [hep-ph]}
  \BibitemShut {NoStop}%
\bibitem [{\citenamefont {Ball}\ \emph {et~al.}(2017)\citenamefont {Ball} \emph
  {et~al.}}]{NNPDF:2017mvq}%
  \BibitemOpen
  \bibfield  {author} {\bibinfo {author} {\bibfnamefont {R.~D.}\ \bibnamefont
  {Ball}} \emph {et~al.} (\bibinfo {collaboration} {NNPDF Collaboration}),\
  }\href {\doibase 10.1140/epjc/s10052-017-5199-5} {\bibfield  {journal}
  {\bibinfo  {journal} {Eur. Phys. J. C}\ }\textbf {\bibinfo {volume} {77}},\
  \bibinfo {pages} {663} (\bibinfo {year} {2017})},\ \Eprint
  {http://arxiv.org/abs/1706.00428} {arXiv:1706.00428 [hep-ph]} \BibitemShut
  {NoStop}%
\bibitem [{\citenamefont {Gao}\ \emph {et~al.}(2022{\natexlab{a}})\citenamefont
  {Gao}, \citenamefont {Hanlon}, \citenamefont {Karthik}, \citenamefont
  {Mukherjee}, \citenamefont {Petreczky}, \citenamefont {Scior}, \citenamefont
  {Syritsyn},\ and\ \citenamefont {Zhao}}]{Gao:2022vyh}%
  \BibitemOpen
  \bibfield  {author} {\bibinfo {author} {\bibfnamefont {X.}~\bibnamefont
  {Gao}}, \bibinfo {author} {\bibfnamefont {A.~D.}\ \bibnamefont {Hanlon}},
  \bibinfo {author} {\bibfnamefont {N.}~\bibnamefont {Karthik}}, \bibinfo
  {author} {\bibfnamefont {S.}~\bibnamefont {Mukherjee}}, \bibinfo {author}
  {\bibfnamefont {P.}~\bibnamefont {Petreczky}}, \bibinfo {author}
  {\bibfnamefont {P.}~\bibnamefont {Scior}}, \bibinfo {author} {\bibfnamefont
  {S.}~\bibnamefont {Syritsyn}}, \ and\ \bibinfo {author} {\bibfnamefont
  {Y.}~\bibnamefont {Zhao}},\ }\href {\doibase 10.1103/PhysRevD.106.074505}
  {\bibfield  {journal} {\bibinfo  {journal} {Phys. Rev. D}\ }\textbf {\bibinfo
  {volume} {106}},\ \bibinfo {pages} {074505} (\bibinfo {year}
  {2022}{\natexlab{a}})},\ \Eprint {http://arxiv.org/abs/2206.04084}
  {arXiv:2206.04084 [hep-lat]} \BibitemShut {NoStop}%
\bibitem [{\citenamefont {Gao}\ \emph {et~al.}(2022{\natexlab{b}})\citenamefont
  {Gao}, \citenamefont {Hanlon}, \citenamefont {Mukherjee}, \citenamefont
  {Petreczky}, \citenamefont {Scior}, \citenamefont {Syritsyn},\ and\
  \citenamefont {Zhao}}]{Gao:2021dbh}%
  \BibitemOpen
  \bibfield  {author} {\bibinfo {author} {\bibfnamefont {X.}~\bibnamefont
  {Gao}}, \bibinfo {author} {\bibfnamefont {A.~D.}\ \bibnamefont {Hanlon}},
  \bibinfo {author} {\bibfnamefont {S.}~\bibnamefont {Mukherjee}}, \bibinfo
  {author} {\bibfnamefont {P.}~\bibnamefont {Petreczky}}, \bibinfo {author}
  {\bibfnamefont {P.}~\bibnamefont {Scior}}, \bibinfo {author} {\bibfnamefont
  {S.}~\bibnamefont {Syritsyn}}, \ and\ \bibinfo {author} {\bibfnamefont
  {Y.}~\bibnamefont {Zhao}},\ }\href {\doibase 10.1103/PhysRevLett.128.142003}
  {\bibfield  {journal} {\bibinfo  {journal} {Phys. Rev. Lett.}\ }\textbf
  {\bibinfo {volume} {128}},\ \bibinfo {pages} {142003} (\bibinfo {year}
  {2022}{\natexlab{b}})},\ \Eprint {http://arxiv.org/abs/2112.02208}
  {arXiv:2112.02208 [hep-lat]} \BibitemShut {NoStop}%
\bibitem [{\citenamefont {Yang}\ \emph {et~al.}(2017)\citenamefont {Yang},
  \citenamefont {Sufian}, \citenamefont {Alexandru}, \citenamefont {Draper},
  \citenamefont {Glatzmaier}, \citenamefont {Liu},\ and\ \citenamefont
  {Zhao}}]{Yang:2016plb}%
  \BibitemOpen
  \bibfield  {author} {\bibinfo {author} {\bibfnamefont {Y.-B.}\ \bibnamefont
  {Yang}}, \bibinfo {author} {\bibfnamefont {R.~S.}\ \bibnamefont {Sufian}},
  \bibinfo {author} {\bibfnamefont {A.}~\bibnamefont {Alexandru}}, \bibinfo
  {author} {\bibfnamefont {T.}~\bibnamefont {Draper}}, \bibinfo {author}
  {\bibfnamefont {M.~J.}\ \bibnamefont {Glatzmaier}}, \bibinfo {author}
  {\bibfnamefont {K.-F.}\ \bibnamefont {Liu}}, \ and\ \bibinfo {author}
  {\bibfnamefont {Y.}~\bibnamefont {Zhao}},\ }\href {\doibase
  10.1103/PhysRevLett.118.102001} {\bibfield  {journal} {\bibinfo  {journal}
  {Phys. Rev. Lett.}\ }\textbf {\bibinfo {volume} {118}},\ \bibinfo {pages}
  {102001} (\bibinfo {year} {2017})},\ \Eprint
  {http://arxiv.org/abs/1609.05937} {arXiv:1609.05937 [hep-ph]} \BibitemShut
  {NoStop}%
\bibitem [{\citenamefont {Ji}\ \emph {et~al.}(2013)\citenamefont {Ji},
  \citenamefont {Zhang},\ and\ \citenamefont {Zhao}}]{Ji:2013fga}%
  \BibitemOpen
  \bibfield  {author} {\bibinfo {author} {\bibfnamefont {X.}~\bibnamefont
  {Ji}}, \bibinfo {author} {\bibfnamefont {J.-H.}\ \bibnamefont {Zhang}}, \
  and\ \bibinfo {author} {\bibfnamefont {Y.}~\bibnamefont {Zhao}},\ }\href
  {\doibase 10.1103/PhysRevLett.111.112002} {\bibfield  {journal} {\bibinfo
  {journal} {Phys. Rev. Lett.}\ }\textbf {\bibinfo {volume} {111}},\ \bibinfo
  {pages} {112002} (\bibinfo {year} {2013})},\ \Eprint
  {http://arxiv.org/abs/1304.6708} {arXiv:1304.6708 [hep-ph]} \BibitemShut
  {NoStop}%
\end{thebibliography}%

\appendix



\end{document}